\documentstyle[prl,aps,multicol,epsfig]{revtex}
\begin{document} 
\draft 
\title{Theory of ARPES intensities from the CuO$_2$ plane}
\author{C. Dahnken and R. Eder}
\address{Institut f\"ur Theoretische Physik, Universit\"at W\"urzburg,
Am Hubland,  97074 W\"urzburg, Germany}
\date{\today}
\maketitle
\begin{abstract}
  We present a theory for the photon energy and polarization dependence
  of ARPES intensities from the CuO$_2$ plane in the framework of strong
  correlation models. We show that for electric field vector
  in the  CuO$_2$ plane the `radiation characteristics' of the 
  $O$ $2p_\sigma$  and $Cu$ $3d_{x^2-y^2}$ orbitals are strongly peaked
  along the CuO$_2$ plane, i.e. most photoelectrons are emitted
  at grazing angles. This suggests that surface states
  play an important role in the observed ARPES spectra, consistent with
  recent data from Sr$_2$CuCl$_2$O$_2$. We show that a combination of
  surface state dispersion and Fano resonance between surface
  state and the continuum of LEED-states may produce a precipitous drop
  in the observed photoelectron current as a function of in-plane
  momentum, which may well mimic a Fermi-surface crossing.
  This effect may explain the simultaneous `observation' of a hole-like and
  an electron-like Fermi surfaces in Bi2212 at different photon energies.
  We show that by suitable choice of photon polarization one can
  on one hand `focus' the radiation characteristics of the in-plane
  orbitals towards the detector and on the other hand make the
  interference between partial waves from different orbitals `more
  constructive'. 
\end{abstract} 
\begin{multicols}{2}
  \section{Introduction}
  Their quasi-2D nature makes cuprate superconductors ideal materials
  for angle resolved photoemission spectroscopy (ARPES) studies, and
  by now a wealth of experimental data is available\cite{ShenDessau}.
  On the other hand, it does not seem as if these data are really
  well-understood, the major reason being that we still lack even a
  rudimentary understanding of the matrix element effects present in
  these materials. It has recently turned out that matrix element
  effects are (or rather: should be) the central issue
  in the discussion of ARPES data.\\
  ARPES is generally believed to measure the single particle spectral
  function, which near the chemical potential $\mu$ (and neglecting
  the finite lifetime) can be written as
  \begin{eqnarray*}
    A(\bbox{k},\omega) &=&
    | \langle \Psi_{QP}(\bbox{k})| c_{\bbox{k},\sigma} | \Psi_0 \rangle |^2\\
    && \Theta(E_{QP}(\bbox{k}) - \mu)\;\delta(\omega-(E_{QP}(\bbox{k}) - \mu))
  \end{eqnarray*}
  Here $E_{QP}$ denotes the dispersion of the `quasiparticle band', $ |
  \Psi_0 \rangle$ and $| \Psi_{QP}(\bbox{k})\rangle$ denote the ground
  state and quasiparticle state, respectively.  In other words, the
  experiment gives a `peak' whose dispersion follows the quasiparticle
  band $E_{QP}(\bbox{k})$, with the total intensity of the peak being
  given by the so-called quasiparticle weight $Z(\bbox{k})= | \langle
  \Psi_{QP}(\bbox{k})| c_{\bbox{k},\sigma} | \Psi_0 \rangle |^2$.  For
  free particles we have $Z(\bbox{k})=1$, whence the only reason for a
  sudden vanishing of the peak with changing $\bbox{k}$ can be the Fermi
  factor $\Theta(E_{QP}(\bbox{k}) - \mu)$, i.e. the crossing of the
  quasiparticle band through the Fermi energy.  Under these
  circumstances, it would be very easy to infer the Fermi surface
  geometry from the measured photoelectron spectra, and indeed this very
  assumption, namely that a sudden drop of the photoemission intensity
  automatically indicates a Fermi level crossing, has long been made in
  the interpretation of all experimental
  spectra on metallic cuprates.\\
  Several experimental findings have shown, however, that this
  assumption is not tenable in the cuprates. The first indication comes
  from the study of the insulating compounds
  Sr$_2$CuCl$_2$O$_2$\cite{Wells} and Ca$_2$CuO$_2$Cl$_2$\cite{Ronning}.
  Although these insulators cannot have any Fermi surface in the usual
  sense, which means that the factor $\Theta(E_{QP}(\bbox{k}) - \mu)$ is
  always equal to unity, the experiments show that also in these
  compounds the quasiparticle peak disappears as one passes from inside
  the noninteracting Fermi surface to outside. Thereby a particularly
  striking feature of the experimental data is the sharpness of the drop
  in spectral weight\cite{Wells}, which is for example along $(1,1)$,
  quite comparable to the drops seen at the `Fermi level crossings' in
  the metallic compounds.  The only possible explanation for this
  phenomenon is a quite dramatic $\bbox{k}$-dependence of the
  quasiparticle weight, $Z(\bbox{k})$. Apparently in these compounds we
  have very nearly $Z(\bbox{k}) \propto \Theta(E_{free}(\bbox{k}) -
  \mu)$, i.e. the $\bbox{k}$ dependence of $Z(\bbox{k})$ resembles
  that of the noninteracting system.\\
  This result immediately raises the question as to how significant the
  `Fermi level crossings' observed in the metallic compounds really are.
  That they may, in some cases, have little or no significance for the
  true Fermi surface topology has been demonstrated by the recent
  controversy as to whether the Fermi surface in the most exhaustively
  studied compound, Bi$_2$Sr$_2$CaCu$_2$O$_{8+\delta}$, is hole-like (as
  inferred from a large number of
  studies\cite{ShenDessau,Golden,Fretwell,Mesot} with photon energy
  $22eV$) or electron-like (as concluded by several recent
  studies\cite{Dessau,Feng,Gromko,Bogdanov} at photon-energy $33eV$).
  Here it should be noted that the true Fermi surface topology is an
  intrinsic property of the material which can under no circumstances
  change with the photon energy.  It follows from these considerations
  that to extract any meaningful information from angle-resolved
  photoemission we need an understanding of the matrix elements and
  their $\bbox{k}$-dependence, as well as other effects which might
  possibly influence the intensity of the ARPES signal.  Motivated by
  these considerations, we have performed a theoretical analysis of the
  spectral weight of strongly correlated electron models.  In section II
  we derive a simple expression for the photoelectron current, which can
  be applied e.g. in numerical calculations for strong correlation
  models. In section III we specialize this to the CuO$_2$ plane, in
  section IV we discuss the angular `radiation characteristics' of a
  Zhang-Rice singlet and how these could be exploited to optimize the
  photoemission intensity from the respective states. We also show that
  photoelectrons are emitted predominantly at small angles with respect to
  the CuO$_2$ plane.
  In section V we point out that this may lead to the injection of these
  photoelectrons into surface resonance states. Fano-resonance between
  the surface resonance and the continuum of LEED states then leads to a
  strong energy dependence of the ARPES signal and we show that already
  a very simple free-electron model can explain the experimental energy
  dependence of the first ionization states in Sr$_2$CuCl$_2$O$_2$ measured
  recently by by D\"urr {\em et al.}\cite{duerr} surprisingly well.  
  In section VI we
  describe how the interplay between surface state dispersion and
  Fano-resonance between the processes of direct emission and emission
  via a surface state can mimic a Fermi level crossing where none
  exists, and suggest that the apparent change in Fermi surface topology
  with photon energy may be due to such `apparent Fermi surfaces'.
  Section VII contains our conclusions.
  \section{Photoemission intensities for strong correlation models}
  Photoelectrons with a kinetic energy in the range $10-100eV$ have
  wavelengths comparable to the distances between individual atoms in
  the CuO$_2$ plane.  It follows that whereas the {\em eigenvalue
    spectrum} of the plane probably can be described well by an
  effective single band model\cite{ZhangRice}, this is not possible for the 
  {\em matrix elements}. We necessarily have to
  discuss (at least) the full three-band model.\\
  Our first goal therefore is to derive a representation of the
  photoemission process in terms of the electron annihilation operators
  for the Cu $3d_{x^2-y^2}$ and O $2p\sigma$ orbitals in the CuO$_2$
  plane. 
  In other words, we seek an operator of the form
  \[
  \tilde{c}_{\bbox{k},\sigma} = m(d_{x^2-y^2}) d_{\bbox{k},\sigma} +
  m(p_x) p_{x,\bbox{k},\sigma} + m(p_y) p_{y,\bbox{k},\sigma}
  \]
  such that the single particle spectral density of this operator
  \[
  \tilde{A}(\bbox{k},\omega) = \frac{1}{\pi} \langle 0|
  \tilde{c}_{\bbox{k},\sigma}^\dagger \frac{1}{\omega - (H-E_0) - i0^+}
  \tilde{c}_{\bbox{k},\sigma}^\dagger|0 \rangle
  \]
  evaluated for the correlated electron model in question reproduces the
  experimental photoemission intensities.  Here $p_{x,\bbox{k},\sigma}$
  and $ p_{y,\bbox{k},\sigma}$ and are the annihilation operators for an
  electron in an $x$- and $y$-directed $\sigma$-bonding oxygen orbital,
  $d_{\bbox{k},\sigma}$ annihilated an electron in the $d_{x^2-y^2}$
  orbital.\\
  To that end, let us first consider the problem of a single atom (which
  may be either Cu or O). 
  The calculation is similar as outlined in Refs. \cite{Matsushita} and
  \cite{Moskvin}.
  We want to study photoionization, i.e. an
  optical transition from a localized valence orbital into a scattering
  state with energy $E$.  The dipole matrix element for light polarized
  along the unit vector $\bbox{\epsilon}$ reads:
  \[
  m_{i\rightarrow f} = \int d\bbox{r}\; \Psi_f^*( \bbox{r})\;(
  \bbox{\epsilon} \cdot \bbox{r})\; \Psi_i( \bbox{r})
  \]
  Here the initial state is taken to be a CEF state with angular
  momentum $l'=1,2$ and crystal-field label
  $\alpha=p_x,p_y,d_{x^2-y^2}\dots$:
  \[
  \Psi_{i}( \bbox{r}) = \sum_{m'} c_{\alpha m'} Y_{l',m'}(\bbox{r}^0)
  \;R_{nl'}(r).
  \]
  We consider this state to be `localized', that means the radial wave
  function $R_{nl'}(r)$ is zero outside the atomic radius $r_0$. For the
  wave function of the final (=scattering) state we choose:
  \[
  \Psi_f( \bbox{r}) = \left\{
    \begin{array}{l c}
      Y_{lm}(\bbox{r}^0)\; \frac{2}{r}\sin(kr+\delta_{l}-\frac{l\pi}{2}) & r > r_0 \\
      \; \\
      Y_{lm}(\bbox{r}^0) \;\frac{1}{n} R_{E,l}(r) & r < r_0
    \end{array} \right.
  \]
  The real functions $R_{nl'}(r)$ and $R_{l}(E)$ both are a solution of
  the radial Schr\"odinger equation for a suitably chosen atomic
  potential. 
  The scattering phase $\delta_{l}$ and
  the prefactor $\frac{1}{n}$ are determined by the condition that the
  wave function $\Psi_f( \bbox{r})$ and its derivative be continuous at
  $r=r_0$.  Details are given in Appendix I.  We note that the
  scattering phase $\delta_{l}$ also plays an important role in the
  interpretation of EXAFS spectra (where it is usually called the
  central atom phase shift) and thus could in principle be determined
  experimentally (although at the relatively high
  photoelectron kinetic energies important for EXAFS are not ideal for ARPES).\\
  Representing the dipole operator as
  \[
  \bbox{\epsilon} \cdot \bbox{r} = \frac{4\pi r}{3} \sum_{\mu=-1}^1
  Y_{1,\mu}^*(\bbox{\epsilon})\; Y_{1,\mu}(\bbox{r}^0),
  \]
  we can rewrite the dipole matrix element as
  \begin{eqnarray}
    m_{i\rightarrow f} &=& \sum_l d_{l,m} \; R_{l,l'}(E) \nonumber \\
    d_{l,m} &=&  \sqrt{ \frac{4\pi}{3} }
    \sum_{m'} \; c_{\alpha m'} \;c^1(l,m;l',m') \;
    Y_{1,m-m'}^*(\bbox{\epsilon}).
  \end{eqnarray}
  Here the radial integral is given by
  \[
  R_{l,l'}(E) = \frac{1}{n} \int_0^{r_0} dr \; r^3\; R_{E,l}(r)\;
  R_{nl'}(r)
  \]
  and the following abbreviation for the angular integrals of three
  spherical harmonics has been introduced:
  \begin{eqnarray*}
    \int d \Omega\;&&
    Y_{lm}^* \;(\bbox{r}^0)\; Y_{1,\mu}(\bbox{r}^0)\;
    Y_{l'm'}(\bbox{r}^0) = \\
    && \;\;\;\;\delta_{m,m'+\mu}\;
    \sqrt{ \frac{3}{4\pi} }\;  c^1(l,m;l',m').
  \end{eqnarray*}
  The constants $c^1(l,m;l',m')$ are well-known in the theory of atomic
  multiplets and tabulated for example in Slater's book\cite{Slater}.
  Knowing the atomic potential the radial integral and thus the entire
  matrix element could now in principle be
  calculated.\\
  Next, we assume that the radial part of $\Psi_f(\bbox{r})$ at large
  distances is decomposed into outgoing and incoming spherical waves.
  Observing the outgoing spherical wave at large distance under a
  direction defined by the polar angles $\Theta_k$ and $\Phi_k$, it may
  locally be approximated by a phase shifted plane wave:
  \[
  \Psi_f( \bbox{r}) \approx \left( \frac{e^{i \bbox{k}\cdot
        \bbox{r}}}{r}\right)\;
  Y_{l,m}(\bbox{k}^0)\;(-i)^{l+1}\;e^{i\delta_{l}},
  \]
  where $\bbox{k} = \sqrt{\frac{2mE}{\hbar^2}} \bbox{r}^0$.\\
  Let us next consider an array of identical atoms in the $(x,y)$-plane
  of our coordinate system (which we take to coincide with the CuO$_2$
  plane in all that follows).  To describe the final state after
  ejection of an electron from the atom at site $j$ we would simply have
  to replace throughout $\bbox{r} \rightarrow \bbox{r} - \bbox{R}_j$ in
  the above calculation, where $\bbox{R}_j$ denotes the position of the
  atom.  If we want to give the created photohole a definite in-plane
  momentum $\bbox{k}_{\|}$, however, we have to form a coherent
  superposition of such states, that means weighted by the Bloch factors
  $\frac{1}{\sqrt{N}}e^{- i \bbox{k}_{\|} \cdot \bbox{R}_j}$, where $N$
  denotes the number of atoms in the plane.  At the remote distance
  $\bbox{r}$ we consequently replace $\bbox{r} \rightarrow \bbox{r} -
  \bbox{R}_j$, leaving the polar angles $\Theta_k$ and $\Phi_k$
  unchanged, multiply by $\frac{1}{\sqrt{N}}e^{- i \bbox{k}_{\|} \cdot
    \bbox{R}_j}$
  and sum over $j$.\\
  The photoelectron wave function at $\bbox{r}$ then becomes
  \begin{eqnarray}
    \Psi_f( \bbox{r})  &\rightarrow&\left( \frac{e^{i \bbox{k}\cdot \bbox{r}} }{r}\right)
    Y_{l,m}(\bbox{k}^0)\;(-i)^{l+1}\;e^{i\delta_{l}}
    \nonumber \\
    &&\;\;\;\;\;\;\;\;
    \frac{1}{\sqrt{N}} \sum_j e^{i (\bbox{k}-\bbox{k}_{\|}) \cdot \bbox{R}_j}
    \nonumber \\
    &=&\left( \frac{e^{i \bbox{k}\cdot \bbox{r}} }{r}\right)
    Y_{l,m}(\bbox{k}^0)\;(-i)^{l+1}\;e^{i\delta_{l}}\nonumber \\
    &&\;\;\;\;\;\;\;\;\;
    \sqrt{N} \delta_{\bbox{k}_{\|} + \bbox{G}_{\|},\bbox{k}_{x,y}}.
  \end{eqnarray}
  Here $\bbox{k}_{x,y}$ denotes the projection of $\bbox{k}$ onto the
  $(x,y)$-plane and $\bbox{G}_{\|}$ is a 2D reciprocal lattice vector.
  An important feature of this result is the fact that by creating a
  photohole with momentum $\bbox{k}_{\|}$ (which must belong to the
  first Brillouin zone) the photoelectrons may well be emitted with the
  3D momentum $(\bbox{k}_{\|}+\bbox{G}_{\|}, k_\perp)$, that means the
  parallel momentum component of the photoelectrons need not
  be equal to $\bbox{k}_{\|}$.\\
  Summing over all possible partial waves $(l,m)$ and introducing the
  abbreviation $\tilde{R}_{l,l'}(E)= R_{l,l'}(E) e^{i\delta_l}$ the
  electron current per solid angle at $\bbox{r}$ due to the in-plane
  orbitals of the type $(l',\alpha)$ finally becomes
  \[
  {\bf j} = \frac{4N \hbar {\bf k}}{m}\; | \sum_{l,m}\; d_{lm}\;
  \tilde{R}_{l,l'}(E)\; Y_{lm}(\bbox{k}^0)\;(-i)^{l+1} |^2 .
  \]
  Here $N$ should be taken equal to the number of unit cells within the
  sample area illuminated by
  the photon beam.\\
  After some algebra (Appendix II) this can be brought to the form
  \begin{equation}
    {\bf j} =  \frac{4N \hbar {\bf k}}{m}\; 
    |\sum_l \tilde{R}_{l,l'}(E)\;(\bbox{v}_{l,\alpha}(\bbox{k}^0) \cdot 
    \bbox{\epsilon})\; |^2.
    \label{sing}
  \end{equation}
  Here all the angular dependence on the shape of the original CEF-level
  has been collected in the vectors $\bbox{v}_{l,\alpha}$ - it is
  important to note, that these vectors are obtained by standard
  angular-momentum recoupling so that the resulting angular dependence
  is {\em exact}.  It is only the `radial matrix elements'
  $\tilde{R}_{l,l'}(E)$ which have to be
  calculated approximately and thus may be prone to inaccuracies.\\
  So far we have limited ourselves to one specific type of orbital 
  in the plane. If we allow photoemission from any type of
  orbital we simply have to add up the prefactors of the
  plane-wave states {\em before} squaring to compute the photocurrent.\\
  Summarizing the preceding discussion we can conclude that the proper
  electron annihilation operator to describe the photoemission process
  would be
  \begin{eqnarray}
  \tilde{c}_{\bbox{k}_{\|},\sigma} &=& \sum_{l,\alpha} 
    e^{ -i \bbox{k}_{\|}\cdot \bbox{R}_\alpha}\;
    \left( \tilde{R}_{l,l'(\alpha)}(E)\;
    ( {\bbox v}_{l,\alpha}(\bbox{k}^0) \cdot
    \bbox{\epsilon})\; \right)\; c_{\alpha,\bbox{k}_{\|},\sigma} 
   \nonumber \\
   &=& \sum_{\alpha} \tilde{\bf v}(\alpha)\cdot
    \bbox{\epsilon}\; c_{\alpha,\bbox{k}_{\|},\sigma} .
  \end{eqnarray}
  Here $\bbox{R}_\alpha$ denotes the position of the orbital $\alpha$ within
  the unit cell.
  Suppressing the spin and momentum index for simplicity, the total
  ARPES intensity then would be given (up to an energy and momentum
  independent prefactor) by
  \begin{eqnarray}
    I(\omega) &=& \sum_{\alpha,\beta} 
  \left(  \tilde{\bf v}(\alpha) \cdot \bbox{\epsilon}\right)^*
    \left( \tilde{\bf v}(\beta) \cdot  \bbox{\epsilon}\right)\;
    \Im R_{\alpha,\beta}(\omega -i 0^+) \\
    \label{intens}
    R_{\alpha,\beta}(z) &=& 
    \frac{1}{\pi}
    \langle 0| c_\alpha^\dagger\; \frac{1}{\omega - (H-E_0) -i0^+}\; c_\beta |0\rangle.
    \label{resolv}
  \end{eqnarray}
  Whereas the scalar products $(\tilde{\bf v}(\alpha) \cdot
  \bbox{\epsilon})$ take into account the interplay between the
  real-space shape of the orbitals, the polarization of the incident
  light and the direction of electron emission, the spectral densities
  $R_{\alpha,\beta}(z)$ incorporate the possible many-body effects in
  the CuO$_2$ planes - we thus have the desired recipe for studying
  photoemission intensities in the framework
  of a strong correlation model. \\
  To conclude this section we note that we have actually performed only
  the first stage of the calculation within the so-called three-step
  model. We give a brief list of complications that we have neglected: 
  the emission from lower planes than
  the first and the extinction of the respective photoelectron
  intensity, any diffraction of the outgoing electron wave function from
  the surrounding atoms, the refraction of the photoelectrons as they
  pass the potential step at the surface of the solid.
  It is thus quite obvious that our theory is strongly simplified.
  \section{Application to the three-band model}
\begin{figure}
    \begin{center}
        \epsfig{file=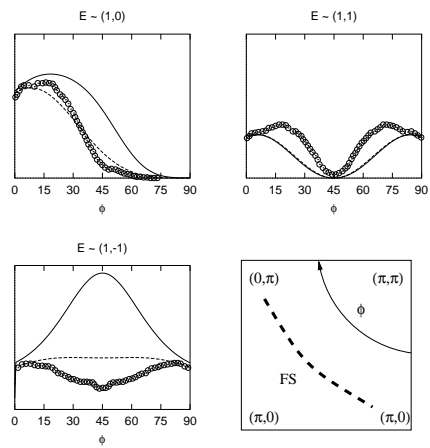,width=6.0cm}
    \end{center}
      \caption[]{
        ARPES intensity at the Fermi 
        energy (in arbitrary units) versus Fermi surface angle
        $\Phi$ (see lower right figure), for three different 
        polarization directions.
        The curves were computed from the tight-binding model 
        (solid line)  and ZRS (dashed line). 
        The photon energy is $h\nu=22\; eV$ and the curves
        are normalized to $\Phi=0$ in panel the upper left panel.
        The points represent the integrated intensities from first 
        the first principle 
        calculation of Bansil and Lindroos
        \cite{banroos}. 
        The lower right panel shows the geometrical details.
        }
      \label{bansilcomp}
  \end{figure}

\begin{figure}[tb]
    \begin{center}
      \epsfig{file=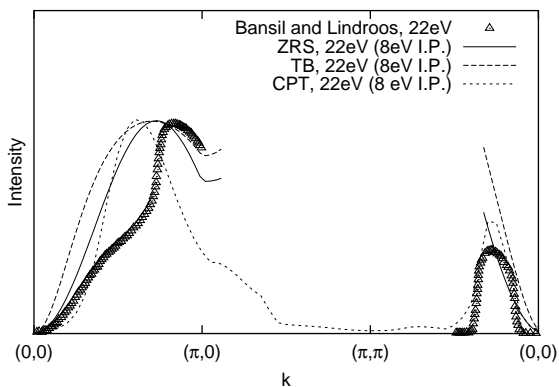}
    \end{center}
      \caption[]{Spectral weight along $(0,0)\rightarrow(\pi,0)$ 
        and $(0,0)\rightarrow(\pi,\pi)$ 
        compared to 
        first principles calculation \cite{banroos} (triangles). 
        ZRS (solid line), 
        tight-binding (long dashes) and 
        CPT (short dashes) method show 
        qualitatively good agreement. 
        The strong decrease of the spectral weight in 
        CPT towards $k=(\pi,0)$ may result 
        from high doping ($\delta=25\%$).}
      \label{bansilcomp2}
  \end{figure}

  We now want to discuss some consequences of the results in the
  preceding section, thereby using mainly the standard three-band
  Hubbard-model to describe the CuO$_2$ plane.  We first consider the
  noninteracting limit $U=0$.  Here we use a $4$-band
  model which was introduced by Andersen {\em et al.}\cite{andersen}
  to describe the LDA
  bandstructure of YBa$_2$Cu$_3$O$_4$. In addition to the 
  Cu 3$d_{x^2-y^2}$ orbital
  and the to $\sigma$-bonding O $2p$ orbitals this model includes a
  Cu $4s$ orbital, which produces the so-called $t'$ and $t''$ terms in the
  single-band model - which are essential to obtain the correct
  Fermi surface topology.  Suppressing the spin index the creation
  operators for single {\em electron} eigenstates read
  \[
  \gamma_{\bbox{k},\nu}^\dagger = \alpha_{1,\nu} p_{\bbox{k},x}^\dagger
  + \alpha_{2,\nu} p_{\bbox{k},y}^\dagger + \alpha_{3,\nu}
  d_{\bbox{k}}^\dagger + \alpha_{4,\nu} s_{\bbox{k}}^\dagger
  \]
  where $\bbox{\alpha}_\nu$ denotes the $\nu^{th}$ eigenvector of the
  matrix
\end{multicols}
\[
H = \left(
  \begin{array}{c c c c}
    0, & 0, & 2i t_{pd} \sin(\frac{k_x a}{2})& 2i t_{sp} \sin(\frac{k_x a}{2})  \\
    0, & 0, & -2i t_{pd} \sin(\frac{k_y a}{2})& 2i t_{sp} \sin(\frac{k_y a}{2})  \\
    - 2i t_{pd} \sin(\frac{k_x a}{2}),& 2i t_{pd} \sin(\frac{k_y a}{2}) & \Delta &0  \\
     -2i t_{sp} \sin(\frac{k_x a}{2})  & -2i t_{sp} \sin(\frac{k_y a}{2})&0 &\Delta_s\\
  \end{array} \right)
\]
\begin{multicols}{2}
  Values of the parameters $t_{pd}$ $t_{sp}$ and $\Delta$ are given in
  Ref. \cite{andersen}.\\
  If we consider the planar momentum
  $\bbox{k}_{\|} = \bbox{G} + \bbox{k}$, where
  $\bbox{k}$ is in the first BZ and
  $\bbox{G}$ is a reciprocal lattice vector,
  the photoemission intensity for the $\nu^{th}$ band becomes
  \begin{eqnarray}
    I &\propto& |
    (\bbox{v}(p_x) \cdot \bbox{\epsilon})\; e^{-i G_x/2} \alpha_1^*
    + (\bbox{v}(p_y) \cdot \bbox{\epsilon})\; e^{-i G_y/2}\alpha_2^*\nonumber \\
    && +  (\bbox{v}(d_{x^2-y^2}) \cdot \bbox{\epsilon})\; \alpha_3^* 
    +  (\bbox{v}(s) \cdot \bbox{\epsilon})\; \alpha_4^* |^2.
    \label{tbmat}
  \end{eqnarray}
  Using this we proceed to a detailed comparison with the work
  of Bansil and Lindroos\cite{banroos}. These authors have
  performed an extensive first-principles study of the
  ARPES intensities in Bi2212, thereby using the more 
  realistic one-step model of photoemission and a complete
  surface band-structure, both for the initial and final state
  wave function. Amongst others, Bansil and Lindroos studied the
  variation of the peak-weight along the Fermi surface
  for given direction of the polarization vector $\bbox{\epsilon}$.
  Their results are shown in Figure \ref{bansilcomp} and compared to the above
  theory.
  Obviously the overall trends seen by Bansil and Lindroos are reproduced
  reasonably well by the present theory and we want to give a brief
  discussion of the mechanisms which lead to these trends.\\
  To begin with, it is to simplest approximation the `oxygen content'
  of the wave function, which determines the 
  ARPES intensity. This is not so much due to the smallness of the
  radial matrix elements for Cu (which are quite comparable to the ones
  for oxygen, see the Appendix), but rather the fact that the 
  partial waves emitted by a $d_{x^2-y^2}$ orbital produce virtually
  no intensity close to the surface normal (as will be shown below). 
  \noindent
  For an electron energy of
  $\approx 20 \;eV$ the detector would have to be placed at 
  $\approx 20^o$ degrees
  from the surface normal (neglecting the refraction by the potential
  step at the surface), and the $d_{x^2-y^2}$ orbital emits practically no
  electrons into this direction.\\
  Next we note that light polarized along the $(1,0)$direction can only 
  excite electrons from $p_x$-type orbitals, 
  light polarized along the $(1,1)$ direction will
  excite the combination $p_x +p_y$, whereas light polarized along
  $(1,-1)$ excites $p_x -p_y$. 
  Finally one has to bear in mind that near $(\pi,0)$ the tight-binding 
  wave function contains only $p_x$-type orbitals (the mixing between
  $p_y(\pi,0)$ and $d_{x^2-y^2}(\pi,0)$ being exactly zero). Therefore
  the polarizations $(1,1)$ and $(1,-1)$ have equal intensity, whereas $(1,0)$
  gives maximum intensity. Near $(0,\pi)$ in the other hand,
  the tight-binding wave function contains only $p_y$,
  whence the polarizations $(1,1)$ and $(1,-1)$ again have
  equal intensity, whereas $(1,0)$ this time gives no intensity.
  For $\bbox{k}\; {\|}\; (1,1)$ exciting 
  $p_x +p_y$ gives no intensity, because this combination does
  not mix with $d_{x^2-y^2}$ and hence is not contained in the
  tight-binding wave function, whereas exciting $p_x -p_y$ 
  gives high intensity. The intensity for
  polarization $(1,0)$ at these momenta is approximately $1/2$ 
  of that for $(\pi,0)$.
  These simple considerations obviously explain the overall
  shape of the curves in Figure \ref{bansilcomp} quite well.
  The main discrepancy between the Bansil-Lindroos theory
  and our calculation is the behaviour of the
  curve for polarization $(1,-1)$. Actually, this is the only curve
  which is not determined by symmetry alone, and subtle details
  of the wave function become important. We believe that the
  parameterization of Andersen {\em et al.} \cite{andersen} 
  does give the correct
  dispersion, but not necessarily 
  the right wave functions.
  Next we consider the intensity variation along 
  various high-symmetry lines in the Brillouin zone, shown in Figure 
  \ref{bansilcomp2}.
  \noindent
  \begin{figure}
    \begin{center}
    \epsfig{file=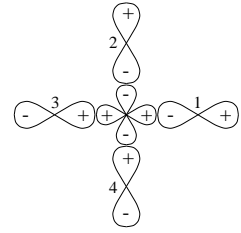,width=3.0cm,angle=0.0}
    \end{center}
    \narrowtext
    \caption[]{Orbitals used for constructing a ZRS and signs of the different
      lobes.}
    \label{plaqu} 
  \end{figure}
  \noindent
  Again, there is reasonable agreement between Bansil and Lindroos and
  our theory.  All in all the comparison shows that also in the results
  of Bansil and Lindroos the
  behaviour of the intensity is to a considerable extent determined
  by the `radiation characteristics' of the orbitals in the CuO$_2$ plane
  and the relative weight of the $p_x$ and $p_y$ orbitals.
  Additional complications like multiple scattering of the photoelectrons,
  Bragg scattering from the BiO surface layer etc. do not seem
  to have a very strong impact on the intensity variations, not even
  at the relatively low photon energy of $22\;eV$.
  Despite its simplicity we therefore believe that our theory 
  has some merit, particularly so because it allows (unlike 
  the single particle calculation of Bansil and Lindroos) to incorporate the 
  effects of strong correlations.
  One weak point of all single-particle-like calculations 
  for the CuO$_2$-plane is the following:
  since (in electron language) the band which forms the Fermi surface
  is the topmost one obtained by mixing
  the energetically higher  Cu 3$d_{x^2-y^2}$-orbital with the
  energetically lower O 2p$\sigma$ orbitals it is clear, that the
  respective wave functions have predominant  Cu 3$d_{x^2-y^2}$ character.
  This is exactly opposite to the actual situation in the cuprates, where the
  fist ionization states in the doped and undoped case are known to
  have predominant O 2p character.
  We now consider the case of large $U$. An exact calculation
  of the single-particle spectra (\ref{resolv}) is no longer possible
  in this case and we have to use various approximations.
  First we study an isolated Zhang-Rice singlet (ZRS) in a single 
  CuO$_4$ plaquette, see Figure \ref{plaqu}.
  The bonding combination of $O$ $2p\sigma$-orbitals is
  \[
  p_{b,\sigma}^\dagger = \frac{1}{2}( p_{1,\sigma}^\dagger
  - p_{2,\sigma}^\dagger - p_{3,\sigma}^\dagger + p_{4,\sigma}^\dagger )
  \]
  and the single-hole basis states (relevant at half-filling) can be written as
  \begin{eqnarray}
    \label{eq:1holestates}
    |1\rangle &=& p_{b,\sigma} |full\rangle, \nonumber \\
    |2\rangle  &=& d_{\sigma} |full\rangle.
  \end{eqnarray}
  Here $|full\rangle$ denotes the $Cu3d^{10} \otimes 4\; O2p^6$ state.
  The single hole ground state then is
  $|\Psi_0^{(1h)}\rangle=\alpha |1\rangle + \beta |2\rangle$
  where $(\alpha,\beta)$ is the
  normalized ground state eigenvector of the matrix
  \begin{equation}
    H = \left( \begin{array}{c c}
        0 & 2t_{pd} \\
        2 t_{pd} & -\Delta
      \end{array} \right)
    \label{singleham}
  \end{equation}
  Two-hole states are obtained by starting from the basis states
  \begin{eqnarray}
    \label{eq:2holestates}
    |1'\rangle &=& p_{b,\uparrow}  p_{b,\downarrow} |full\rangle \nonumber \\
    |2'\rangle &=& \frac{1}{\sqrt{2}}( p_{b,\uparrow}  d_{\downarrow} +
    d_{\uparrow}  p_{b,\downarrow})  |full\rangle \nonumber \\
    |3'\rangle &=& d_{\uparrow}  d_{\downarrow} |full\rangle
  \end{eqnarray}
  The two-hole ground state of the plaquette reads
  $|\Psi_0^{(2h)}\rangle=\alpha' |1'\rangle + \beta' |2'\rangle + \gamma' |3'\rangle$, where
  $(\alpha',\beta',\gamma')$ is an eigenvector of
  \begin{equation}
    H = \left( \begin{array}{c c c}
        0                 &  2\sqrt{2}t_{pd} & 0\\
        2\sqrt{2} t_{pd} & -\Delta          & 2\sqrt{2}t_{pd} \\
        0                 &  2\sqrt{2}t_{pd} &  -2\Delta +U
      \end{array} \right).
    \label{doubleham}
  \end{equation}
  If we want to make contact with the $t-J$ model, where
  a `hole' at site $i$ stands for a ZRS in the plaquette centered on the
  copper site $i$, we have to
  incorporate a phase factor of $e^{-i \bbox{k}_{\|} \cdot \bbox{R}_i}$,
  into the definition of $|\Psi_0^{(2h)}\rangle$,
  where $\bbox{R}_i$ is the position of the central Cu orbital.
  In other words, the matrix element for the creation of a `hole'
  in the $t-J$ model is
  \begin{equation}
  m_{ZRS} = e^{i \bbox{k}_{\|} \cdot \bbox{R}_i}
  \langle \Psi_0^{(2h)} | \tilde{c}_{\bbox{k},\sigma} |\Psi_0^{(1h)}\rangle .
  \label{mzrs}
  \end{equation}
  The matrix elements for the creation of
  $e^{-i \bbox{k}_{\|} \cdot \bbox{R}_j} p_{b,\sigma}|full\rangle$ and
  $e^{-i \bbox{k}_{\|} \cdot \bbox{R}_j} d_{\sigma}|full\rangle$  are
  \begin{eqnarray}
    m_b &=& i \left( \bbox{v}(p_y) \sin(\frac{k_y}{2}) -
      \bbox{v}(p_x) \sin(\frac{k_x}{2} )\right)\cdot \bbox{\epsilon},\nonumber \\
    m_d &=& \bbox{v}(d_{x^2-y^2}) \cdot \bbox{\epsilon}.
    \label{ZRSform}
  \end{eqnarray}
  Finally, the matrix element for creation of a ZRS from the
  single-hole ground state becomes
  \begin{eqnarray*}
    m_{ZRS} &=& \alpha'^* \alpha m_b + \frac{\beta'^*}{\sqrt{2}}(
    \alpha m_d + \beta m_b ) \\
    && + \gamma'^* \beta m_d.
  \end{eqnarray*}
  Using this expression, the intensity expected for a ZRS with momentum 
  $\bbox{k}$ can be calculated as a function of photon polarization and
  energy.  This will be discussed in the next sections.\\
  \begin{figure}
    \begin{center}
    \epsfxsize=6.0cm
    \epsfig{file=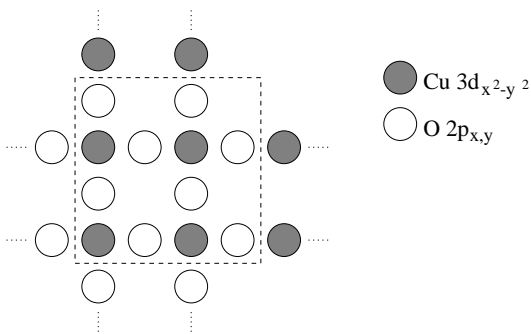,width=7.0cm,angle=0.0}      
    \end{center}
    \narrowtext
    \caption[]{Three band Hubbard cluster used
      in the Lanczos diagonalization. The hopping
      across the cluster boundary (dashed line) is
      treated to lowest order strong coupling
      perturbation theory.}
    \label{fig1}
  \end{figure}
  \noindent
  In comparing the intensities calculated from our above theory 
  for a single ZRS to
  experiment one would be implicitly assuming that the first ionization states
  can be described completely by a coherent superposition
  of ZRS {\em in a single plaquette}.
This need not
be the case and in order to get at least a rough feeling for the effects
of
`embedding the plaquette in a lattice' we use the technique of
cluster perturbation theory (CPT) to study the extended system.
CPT, which was first suggested by Senechal {\em et al.}\cite{Senechal},
is a technique to `extrapolate' the single-electron's
Green's function calculated on a finite cluster to an infinite periodic
system. It is based on an perturbative treatment of the intercluster hopping.
The cluster we used for the present calculation
is a quadratic arrangement of four
unit cells each containing one Cu d$_{x^2-y^2}$ and two O p$_{x,y}$
orbitals as shown in figure \ref{fig1}. The Green's function
\[
G_{ij\sigma}=\left< \Phi_0\right|
c_{i\sigma}\frac{1}{z-H}c^\dagger_{j\sigma}\left| \Phi_0\right>
+\left< \Phi_0\right| c^\dagger_{j\sigma}\frac{1}{z-H} c_{i\sigma}\left|
\Phi_0\right>
\]
  of this cluster with open boundary conditions
  is calculated by the Lanczos method. The boundary orbitals of the
  cluster
  are connected to the adjacent cluster by the Fourier transform of
  the intercluster hopping $V({\bf Q})$,
  where ${\bf Q}$ is a ``superlattice'' wave-vector restricted to the
  smaller Brillouin zone formed by the now enlarged lattice of the
  clusters.
  The perturbative treatment of the intercluster hopping
  in lowest order yields an RPA like expression for the
  approximate Green's function.
  \[
  {\cal G}_{ij\sigma}\left( {\bf Q},\omega\right)=\left(
\frac{1}{G^{-1}(\omega)_{\sigma}-V({\bf Q})}\right)_{ij}
.
  \]
  This mixed representation can then be transformed into momentum space
  by a Fourier transformation thereby taking into account
  the geometry of the cluster:
  \[
  G_{CPT}\left({\bf k},\omega\right)= \frac{1}{N} \sum_{ij}
  e^{i{\bf k}\left( {\bf r}_i-{\bf r}_j\right)}
  {\cal G}_{ij\sigma}\left( {\bf N}{\bf k},\omega\right).
  \]
  This result is exact for $U=0$ and has been shown\cite{Senechal}
  to produce good results also for large and
  intermediate $U$. In the following sections
  we will use this technique for the
  approximate calculation of the single-particle spectral densities
  as a valuable cross-check for the
  calculations based on a single ZRS. While CPT still is far from
  being rigorous, it includes some effects which result
  from embedding the ZRS in a lattice and, as we will in fact see,
  it always gives results which are very similar to those
  for a ZRS.\\
  To conclude this section we want to give a simple estimate
  for the photoelectron kinetic energy $T$ as a function of photon energy
  $h\nu$.
  It is easy to see that in the absence of any potential step at the
  surface of the solid we would have
  \begin{equation}
    T = h\nu + E_0^{(N)} - E_{\bbox{k}}^{(N-1)}
  \end{equation}
  where $E_0^{(N)}$ is the energy of the ground state of the
  solid with $N$ electrons
  and $E_{\bbox{k}}^{(N-1)}$ the final state of the solid.
  We approximate 
  \begin{equation}
    E_0^{(N)} - E_{\bbox{k}}^{(N-1)} \approx
    ( E_0^{1h} - E_0^{2h} ) - E_{2p}
  \end{equation}
  Here $E_0^{nh}$ are the ground state energies of a CuO$_4$
  plaquette with $n$ holes, obtained by diagonalizing the matrices 
  (\ref{singleham}) and (\ref{doubleham}). 
  In these matrices the energy of the $O2p$ level, $E_{2p}$, has been
  chosen as the zero of energy, whence it has to be taken
  into account separately.
  For the 
  `standard values' of the parameters $t_{pd}=-1.3\;eV$, $\Delta=3.6\;eV$ and 
  $U=10.5\;eV$ this
  gives $E_0^{1h} - E_0^{2h} = 1.93\;eV$.
  Using the estimate $E_{2p} \approx -10\;eV$ from our atomic LDA calculation we find
  \begin{equation}
    T \approx h\nu - 8\; eV,
  \end{equation}
  which we will use in all that follows.\\
  \begin{figure}[htbp]
    \begin{center}
      \epsfig{file=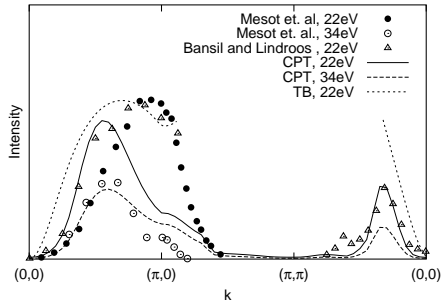,width=6.0cm}
    \end{center}
    \caption[]{Spectral weight along $(0,0)\rightarrow(\pi,0)$ and 
      $(0,0)\rightarrow(\pi,\pi)$
      compared to experimental data from references \cite{banroos,Mesot}. }
    \label{bansilcomp4}
  \end{figure}
  \noindent
  As a first application, Figure \ref{bansilcomp4} then
  shows the intensity of the 
  topmost   ARPES peak calculated by CPT and compares this to
  experimental data. Reasonable qualitative
  agreement can be found, especially for the CPT calculation, 
  although the shift of the spectral weight maximum towards $k=(\pi,0)$ for
  $22\;eV$ is not
  reproduced. The much better agreement for $E_{photon}=34eV$ suggests that 
  for the lower photon energy of $22\;eV$ additional effects (such as multiple
  scattering corrections or Bragg-scattering from the BiO top-layer)
  are more important.
  \section{Application to experiment}
\begin{figure}
    \begin{center}
    \epsfig{file=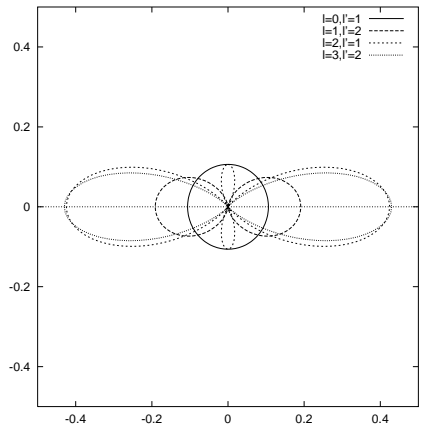,width=6.0cm,angle=0.0}
    \end{center}
    \epsfxsize=6.0cm
    \narrowtext
    \caption[]{`Radiation characteristics'  
      $|\bbox{v}_{l,l'}\cdot\bbox{\epsilon}|^2$ for the different
      partial waves from
      a ZRS with momentum $(\pi,0)$. The figure shows the
      $\Theta$-dependence of the respective partial wave within
      in the $x-z$ plane. 
      The polarization vector $\bbox{\epsilon}$ of the exciting light
      is assumed to be in $x$-direction.}
    \label{char} 
  \end{figure}

  Coming back to the theory for the ZRS we can already draw
  conclusions of some importance.
  By combining the expressions for the vectors $\bbox{v}_{l,\alpha}$
  from  Appendix II 
  with the `form factor' of a ZRS, (\ref{ZRSform}), the
  expression for the photocurrent can be brought to the form
  \begin{equation}
  \bbox{j} \propto | \sum_{l,l'} \tilde{R}_{l,l'}(E)
  \;(\bbox{v}_{l,l'}(\bbox{k}_{\|},\bbox{k})\cdot{\bf \epsilon})\; |^2.
  \label{total}
  \end{equation}
  We remember that $\bbox{k}_{\|}$ is the momentum of the photohole
  (which is within the first Brillouin zone of the
  CuO$_2$ plane), whereas $\bbox{k}$ is the momentum of the
  escaping photoelectron.
  Using the expressions from Appendix II it is now a matter of
  straightforward algebra to derive the following expressions
  for $\bbox{k}_{\|}=(k,0)$, and the
  `optimal' photon polarization for momenta along $(1,0)$,
  $\bbox{\epsilon}=(1,0,0)$:
  \begin{eqnarray}
    \bbox{v}_{0,1}\cdot\bbox{\epsilon} &=& \frac{\sin(k/2)}{\sqrt{12\pi}} ,\\
    \bbox{v}_{1,2}\cdot\bbox{\epsilon} &=& -i \sqrt{\frac{3}{20\pi}} \;\sin(\Theta),\\ 
    \bbox{v}_{2,1}\cdot\bbox{\epsilon} &=& \frac{\sin(k/2)}{\sqrt{12\pi}}
    \left( \frac{3}{2}\cos(2\Theta) -\frac{1}{2}\right),\\
    \bbox{v}_{3,2}\cdot\bbox{\epsilon} &=& -i \sqrt{\frac{3}{20\pi}}\; \sin(\Theta)\; 
     (\frac{5}{4}\cos(2\Theta) - \frac{1}{4}).
  \end{eqnarray}
  Figure \ref{char} shows the $\Theta$-dependence
  of these expressions.
  \noindent
  Obviously a ZRS with momentum near $(\pi,0)$ emits photoelectrons 
  predominantly 
  parallel to the CuO$_2$ plane if it is excited with
  light polarized along $(1,0)$
  within the CuO$_2$ plane. The situation is similar,
  though not as pronounced, for other momenta, such as 
  $(\frac{\pi}{2},\frac{\pi}{2})$. It should be noted, that the
  vectors $\bbox{v}_{l,l'}(\bbox{k}_{\|},\bbox{k})$ are computed {\em exactly}
  namely by elementary angular-momentum recoupling. The fact that
  a ZRS with momentum near $(\pi,0)$ emits photoelectrons predominantly 
  at small angles with respect to the CuO$_2$ plane therefore
  is a rigorous result.\\
  The radial matrix elements $\tilde{R}_{l,l'}(E)$ which also enter in
  (\ref{total}) actually tend to suppress the
  emission close to the surface normal even more. 
  Namely one finds (Appendix I) that
  $R_{0,1} \approx 0.2\; R_{2,1}$, that means the $s$-like
  partial wave (which would contribute strongly to emission
  at near perpendicular directions, see Figure \ref{char}) 
  has a very small weight due to the radial matrix elements.
  We note in passing that the smallness of the ratio $R_{0,1}/R_{2,1}$
  is well-known in the EXAFS literature\cite{exafs}.\\
  Combining the partial waves in Figure \ref{char} with the proper
  radial matrix elements $\tilde{R}_{l,l'}(E)$ as in (\ref{total})
  we expect to obtain a
  curve with a minimum for some finite $\Theta$ (mainly due to to the
  node in the dominant $d$-like $l=2$ partial wave emitted by the
  2p $l'=1$ orbital (see Figure \ref{char}). This may in fact explain
  a well-known\cite{Dessau} effect in ARPES, namely the
  the relatively strong asymmetry with respect to $(\pi,0)$ of the ARPES intensity for
  $\bbox{k}_{\|}$ along $(1,0)$ which 
  \begin{figure}
    \begin{center}
    \epsfig{file=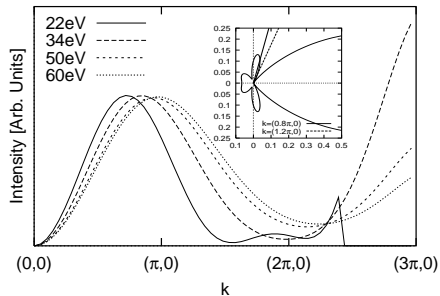,width=6.0cm,angle=0.0}
    \end{center}
    \narrowtext
    \caption[]{ARPES intensity as a function of momentum along $(1,0)$
     for different photon energies, calculated by for an isolated ZRS.
     The inset shows the radiation characteristics of a ZRS at $(\pi,0)$
     and the directions at which a detector would have to be placed at photon energy
     $22\;eV$ to observe the respective momentum. Thereby the
     work function is neglected.}
    \label{ZRS-Along_1_0_comb.eps}
  \end{figure}
  \noindent
  is seen at $22\;eV$ photon energy
  but not at $34\;eV$. Figure \ref{ZRS-Along_1_0_comb.eps} shows
  the calculated intensities as a function of in-plane momentum.
  For $h\nu \approx 22\;eV$ the angles$\Theta$ which would be
  appropriate to observe $\bbox{k}_{\|}$ slightly beyond $(\pi,0)$
    (i.e. in the second zone) are such, that one is `looking into the node' 
  of the radiation characteristics of the ZRS, whereas
  for smaller $\bbox{k}_{\|}$ one is looking at the maximum (see the inset).
  Hence there is a strong
  asymmetry around $(\pi,0)$. Increasing the energy to $\ge 30 eV$
  the angles $\Theta$ becomes smaller, one is no longer sampling
  the node and the the intensity is much more symmetric around $(\pi,0)$.\\
  As shown in the preceding discussion, our theory reproduced, despite its simplicity,
  some experimental features seen in the cuprates.
  We therefore proceed to address potential applications in experiment. 
  Thereby the main goal is
  to find experimental conditions under which the ZRS-derived state at a given 
  momentum ${\bf k}_{\|}$ can be observed with the highest intensity.
  To that end, we can vary different experimental parameters, mainly the direction
  of the photon polarization and the Brillouin zone in which we are measuring.\\
  We first consider the case of normal incidence of the light,
  that means the electric field vector $\bbox{\epsilon}$ is in the CuO$_2$ plane.
  Rotating $\bbox{\epsilon}$ in the plane will change
  the intensity from the ZRS-derived states in a systematic way, and 
  for some angle $\Phi_E$ it will be maximum.
  Along high symmetry directions like $(1,1)$ or $(1,0)$ the optimal
  polarization can be deduced by symmetry considerations, because
  the ZRS has a definite parity under reflections by these directions.
  For momenta $\bbox{k}$ along $(1,1)$ $\bbox{\epsilon}$
  has to be perpendicular to $\bbox{k}$, whereas along $(1,0)$ it 
  must be parallel\cite{ShenDessau}.
  For nonsymmetric momenta, however, this symmetry analysis is not
  possible and one has to calculate the optimum direction.
  We note that apart from merely enhancing the intensity of the ZRS, 
  knowledge of the polarization dependence of the ARPES intensity
  of the `ideal' ZRS would allow also to separate the original
  signal from ZRS-derived 
  states from any `background', that means photoelectrons which have undergone
  inelastic scattering  on their way to
  the analyzer. Since it is plausible that these background electrons have
  more or less lost the information about the polarization of the
  incoming light, their contribution should be insensitive
  to polarization. Taking the spectrum at the `optimal angle'
  for the ZRS and perpendicular to it then would allow to remove
  the background by merely subtracting the two spectra
  (provided one can measure the absolute intensity). Along the high
  symmetry directions $(1,0)$ and $(1,1)$ the feasibility of this
  procedure has recently been demonstrated by Manzke {\em et al.} \cite{Manzke}
  and using the calculated optimal polarizations this analysis could
  be extended to any point in the BZ.\\
  For the special case of the ZRS, and neglecting the contribution of
  copper altogether, it is possible to give a rather simple
  expression for the optimal angle.
  The matrix element for creating the bonding combination
  of O 2p$\sigma$ orbitals is
  \begin{eqnarray}
    m_{ZRS} &\propto& i
    \left(\sin(\Phi_E)\sin(\frac{k_y}{2})-\cos(\Phi_E)\sin(\frac{k_x}{2})\right).
  \end{eqnarray}
  We thus find the angles which give minimum and maximum intensity:
  \begin{eqnarray}
    \sin(\Phi_{E,min})\sin(\frac{k_y}{2})-\cos(\Phi_{\min})\sin(\frac{k_x}{2})
    &=& 0 \nonumber \\
    \cos(\Phi_{E,max})\sin(\frac{k_y}{2})+\sin(\Phi_{\max})\sin(\frac{k_x}{2})
    &=& 0 
  \end{eqnarray}
  We see that always $\Phi_{E,min} = \Phi_{E,max} + \frac{\pi}{2}$.
  Moreover, in this approximation the intensity is exactly zero for 
  $\Phi_{E,min}$, and the angle for maximum intensity is
  \begin{equation}
    \Phi_{E,max} = - arctan\left( 
      \frac{\sin(\frac{k_y}{2})}{\sin(\frac{k_x}{2})} \right).
    \label{ZRSsimp}
  \end{equation}
  This expression takes a particularly simple form along the line
  $(\pi,0) \rightarrow (0,\pi)$, where 
  $\Phi_{\max} = -k_y/2$. This reproduces the known values
  $\Phi_{\max} = 0$ along $(1,0)$, $\Phi_{\max} = -\frac{\pi}{4}$ along $(1,1)$
  and  $\Phi_{\max} = -\frac{\pi}{2}$ along $(0,1)$.
  Despite its simplicity formula (\ref{ZRSsimp})
  gives a quite good estimate for
  the optimum polarization.
  Numerical evaluation shows that the lines of constant
  $\Phi_{\max} $ in $\bbox{k}$-space are to very good approximation
  straight lines through the
  center of the Brillouin zone.
  This means that in practice one could adjust the polarization angle
  $\Phi_E$ once to
  either $\Phi_{E,max}$ or $\Phi_{E,min}$, and then scan an entire straight line
  through the origin by varying the emission angle 
  $\Theta_k$ without having to change the polarization along the way.
  \begin{figure}
    \begin{center}
    \epsfig{file=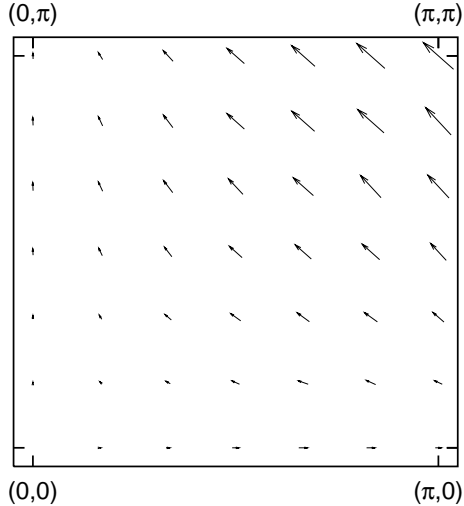,width=7.0cm,angle=0.0}      
    \end{center}
    \narrowtext
    \caption[]{Optimal in-plane polarization for observing the
      ZRS at the respective position in the Brillouin zone
      by ARPES with photoelectron kinetic energy  energy. The calculation is
      done for an isolated ZRS.}
    \label{max-zrs} 
  \end{figure}
  \noindent
  Figures \ref{max-zrs} and \ref{max-cpt} then show the optimal angle for observation
  of the first ionization state 
  for momenta in the first
  Brillouin zone. For perpendicular polarization the intensity
  practically vanishes.
  In the Figure we compare the `isolated ZRS' 
  and the `embedded ZRS' whose spectra are
  obtained by CPT.  Both methods of calculation give very similar
  results for the optimal angle, which in turn agrees very well with the
  simple estimate \ref{ZRSsimp}.
  Since along the high symmetry lines
  $(1,0)$ and $(1,1)$ the optimal $\Phi_E$ is determined by symmetry alone
  there is obviously not so much freedom to `interpolate' smoothly
  between these values. \\
  Next, moving to a higher Brillouin zone allows to enhance the intensity
  of the ZRS, as can be seen from Table \ref{table1} and \ref{table2}.
  Table \ref{table1} shows the ratio $I/I_0$ 
  of the intensity obtainable by measuring
  the ZRS at $(\frac{\pi}{2},\frac{\pi}{2})$ at the `original position'
  (there the polarization has to be perpendicular to (1,1) by symmetry)
  and the intensity that can be obtained by measuring in a
  higher zone with optimized polarization, both within in the CuO$_2$ plane
  and, for later reference, also out of plane.  
  \begin{figure}
    \begin{center}
      \epsfig{file=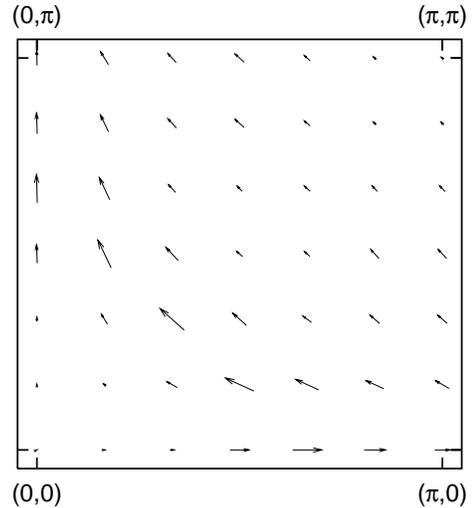,width=7.0cm,angle=0.0}
    \end{center}
    \narrowtext
    \caption[]{Optimal in-plane polarization for observing the
      first ionization state of the correlated CuO$_2$ plane.
      The intensities are calculated by integrating the spectral weight
      obtained by CPT within 300meV.}
    \label{max-cpt} 
  \end{figure}
  \noindent
  The same information
  is given for $(\pi,0)$ in table \ref{table2}. Obviously, an enhancement
  of a factor of $2$ or more can be achieved by proper choice of the
  experimental conditions. In view of the large `background'
  in the experimental spectra, even such a moderate enhancement may be
  quite important.\\
  We note that the physical origin of the variation
  of intensity with $\Phi_E$ and order of the Brillouin zone 
  \begin{figure}
    \begin{center}
      \epsfig{file=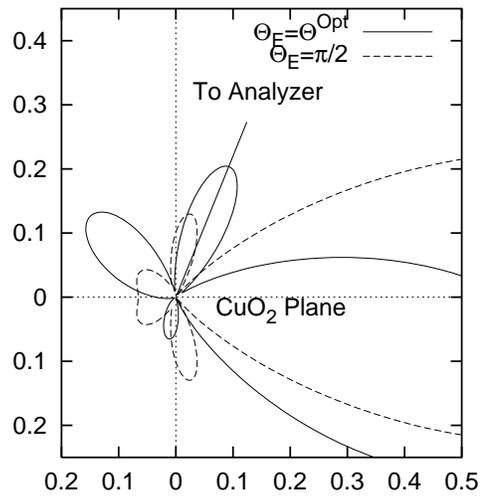,width=7.0cm}      
    \end{center}
    \narrowtext
    \caption[]{`Radiation characteristics' within the $(x-z)$ plane
      of a ZRS excited with light polarized in the $(x-z)$ plane.
      The polar angles are $\Theta_E=\frac{\pi}{2}$ (dashed line)
      and $\Theta_E=\Theta_{E,opt}\approx\frac{\pi}{3}$ (full line). 
      The skew line shows the
      direction where the detector would have to be placed for the photon 
      energy of $22\;eV$ (work function neglected).}
    \label{focus2} 
  \end{figure}
  \noindent
  is the interference between
  the processes of creating a photohole in $p_x$-like and $p_y$-like
  $O$ $2p\sigma$ orbitals. 
  By changing $\Phi_E$ the relative intensity and phase
  between photoholes in the two types of orbitals can be
  tuned. Thus, one can always find an angle $\Phi_{E,max}$ where the
  interference between these two types of orbital
  is maximally constructive. This angle depends on
  the relative phase between the two orbitals in the wave function
  and thus is specific for the state in question.
  Next, we study a quite different effect, namely what happens if we
  tilt the polarization vector $\bbox{\epsilon}$ out of the
  CuO$_2$ plane. To illustrate the usefulness
  of this, we again consider a momentum along the high symmetry line
  $(0,0)\rightarrow (\pi,0)$ and assume that the polarization
  vector $\bbox{\epsilon}$ is within in $x-z$ plane
  It follows from symmetry
  considerations that any component of the light
  perpendicular to this plane (`s-polarization') cannot
  excite any ZRS-derived states.
  In other words we assume that  $\Phi_E$ but that
  $\Theta_E$ is variable. The contribution
  from the bonding combination of $O$ $2p\sigma$ orbitals then becomes
\end{multicols}
\begin{eqnarray*}
  m_{ZRS}&=& -i \xi_1 \Bigg[ 
  \left(
    \frac{1}{\sqrt{3}} \tilde R_{01}s
    +\frac{1}{\sqrt{15}} \tilde R_{21}d_{3z^2-r^2}
    -\frac{1}{\sqrt{5}} \tilde R_{21}d_{x^2-y^2}    
  \right) \sin(\Theta)  
  +\frac{1}{\sqrt{5}} \tilde R_{21}d_{xz}\cos(\Theta)
  \Bigg] \cr
  &&+i\xi_2 \Bigg[
  \left( 
    -\frac{1}{\sqrt{5}} \tilde R_{321} p_x
    -\frac{1}{\sqrt{70}} \tilde R_{323} f_{x\left( 5z^2-1\right)}
    +\sqrt{\frac{3}{70}} \tilde R_{323} f_{ x^3-3xy^2}
  \right)\sin(\Theta)
  +\frac{1}{\sqrt{7}} \tilde R_{323} f_{z\left( x^2-y^2\right)} \cos(\Theta) 
  \Bigg], \cr
\end{eqnarray*}
\widetext
\begin{multicols}{2}
  where $\xi_1=\alpha'^* \alpha+\frac{\beta'^* \beta}{\sqrt{2}}$ and 
  $\xi_2=\gamma'^* \beta+\frac{\beta'^* \alpha}{\sqrt{2}}$. 
  Tilting the electric field vector out of the plane thus admixes
  the $d_{xz}$ harmonic into the radiation characteristics, which has its maximum 
  intensity at an angle of $45^o$ with respect to the CuO$_2$ plane.
  Clearly, this enhances the intensity at intermediate $\Theta_k$,
  particularly so if constructive interference with
  the $d_{x^2-y^2}$ and $d_{3z^2-r^2}$ harmonics (which are
  excited by the $x$-component of $\bbox{\epsilon}$) occurs.
  Since the $d_{xz}$ on one hand and the $d_{x^2-y^2}$ and $d_{3z^2-r^2}$ on
  the other have opposite parity under reflection by the $y-z$ plane,
  constructive interference for some momentum $(k,0)$ automatically implies
  destructive interference for momentum $(-k,0)$. Tilting the
  electric field out of the plane thus amounts to `focusing'
  the photoelectrons towards the detector.
  This effect is illustrated in Figure \ref{focus2}, which shows the 
  angular variation within the $(x-z)$ plane of
  the photocurrent radiated by a ZRS.
  \begin{figure}
    \begin{center}
      \epsfig{file=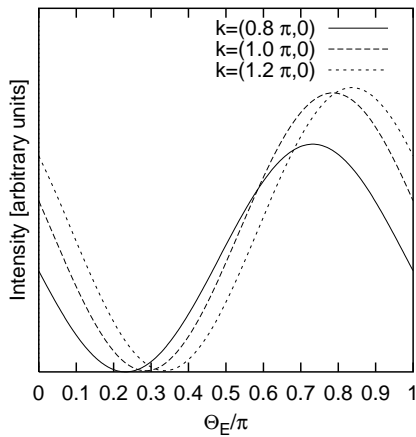,width=6.0cm}      
    \end{center}
    \narrowtext
    \caption[]{Intensity of ZRS derived states
      with different momenta along the $(1,0)$-direction, 
      as a function of the polar angle $\Theta_E$.
      The light is polarized in the $(x-z)$ plane, i.e. $\Phi_E=0$,
      the photon energy is $22\;eV$ (work function neglected).}
    \label{thetascan} 
  \end{figure}
  \noindent
  Figure \ref{thetascan} then shows the intensity as a function of
  $\Theta_E$ for various momenta along $(0,0)\rightarrow (\pi,0)$.
  Remarkably enough, the enhancement of current towards the detector may be up 
  to a factor of $3$ as compared to polarization in the plane.
  Exploiting this effect could enhance the photoemission intensity considerably
  in the region around $(\pi,0)$ - which might be important, 
  because this is precisely
  the `controversial' region in $\bbox{k}$-space which makes the difference
  between hole-like\cite{ShenDessau,Golden,Fretwell,Mesot}
  and electron-like\cite{Dessau,Feng,Gromko,Bogdanov} Fermi surface in
  Bi2212 and where bilayer-splitting\cite{Armitage,Chuang} 
  should be observable.
  One {\em caveat} is the fact, that for a polarization which is not in the
  plane one has to compute the actual field direction by use of the 
  Fresnel formulae\cite{Matzdorf}. The value of the dielectric constant, 
  however, which has
  to be chosen for such a calculation, is very close to $1$.
  Tables I and II also give the optimal angles $\Theta_E$ and $\Phi_E$
  for observing $(\frac{\pi}{2},\frac{\pi}{2})$ and $(\pi,0)$ in
  higher Brillouin zones. Obviously, by choosing the right zone and
  the right polarization a considerable enhancement of the intensity can be
  achieved.
\section{Surface resonances and the energy dependence of the intensity}
\begin{figure}
  \begin{center}
    \epsfig{file=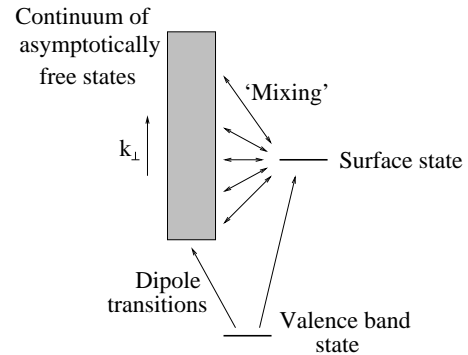,width=7.0cm,angle=0.0}    
  \end{center}
\narrowtext
\caption{Energy level diagram for ARPES via a surface state.}
\label{fano}
\end{figure}

We have seen in the preceding sections that the orbitals
from which the ZRS is built emit photoelectrons
predominantly at grazing angles with respect to the CuO$_2$ plane
(see Figure \ref{char}). In most cases electrons emitted with momentum
$(\bbox{k}_{\|}+\bbox{G}_{\|},k_\perp')$ will be simply `lost'
because they will not reach a detector positioned
to collect electrons with momentum
$(\bbox{k}_{\|},k_\perp)$ (it might happen that the
 photoelectron `gets rid of its $\bbox{G}_{\|}$' by Bragg-scattering
at the BiO surface layer - a possibility which we are neglecting here). 
It may happen, however,
that the motion with momentum $\bbox{k}_{\|}+\bbox{G}_{\|}$
parallel to the surface `consumes so much energy'
that there is hardly any energy left for the motion perpendicular
to the surface. This will happen whenever the kinetic
energy $T$ is just above the so-called emergence
condition for the reciprocal lattice vector $\bbox{G}_{\|}$:
\begin{equation}
T = \frac{\hbar^2}{2m}(\bbox{k}_{\|} + \bbox{G}_{\|})^2.
\label{emergence}
\end{equation}
In other words: the kinetic energy is `just about
sufficient' to create a free electron state with 3D momentum
$(\bbox{k}_{\|} + \bbox{G},0)$. In this case the 
energy available for motion perpendicular to the surface is not sufficient
for the photoelectron to
surmount the energy barrier at the surface and it
will be reflected back in to the solid. \\
On the other hand, if the `perpendicular energy' is below the
energy of the Hubbard gap as well, 
the electron cannot re-enter the solid either,
because there are no single-particle states available in the solid
with the proper energy. The situation thus is quite analogous
to the case of the so-called Shockley state seen on the
Cu surface: a combination of surface potential and
a gap in the single-particle DOS causes the electron to be trapped
at the surface of the solid.
For an extensive review of the properties of such so-called
surface resonances see Ref.\cite{McRae}. \\
The existence of such surface resonances also in a cuprate-related material
has been established by Pothuizen\cite{hans_thesis}.
In an extensive EELS study of Sr$_2$CuCl$_2$O$_2$, Pothuizen
could in fact identify not just one, but a total of
$3$ such states (in some cases with clearly resolved dispersion)
in the energy window $0-30\;eV$.
His results show very clearly that such surface states do exist
in Sr$_2$CuCl$_2$O$_2$ - whether this holds true for other
cuprate materials has not been established yet, but
for the moment we will take their existence for granted.\\
To proceed in at least a semiquantitative
way we use a very simple approximation\cite{McRae}. 
We decompose the potential felt by an electron at the surface as
\begin{eqnarray}
V(\bbox{r}) &=& V_{av}(z) + V_1(\bbox{r}),\\
 V_{av}(z) &=& \frac{1}{a^2}\int_{0}^a dx \int_{0}^a dy\; V(x,y,z),\\
V_1(\bbox{r}) &=& V(\bbox{r}) -  V_{av}(z),
\end{eqnarray}
($a$ denotes the planar lattice constant)
and assume that $V_1$ can be treated as a perturbation.
The eigenstates of an electron in the potential $ V_{av}(z)$ can be
factorized:
\[
\Psi_{\bbox{G}_{\|},\mu}(\bbox{r}) = e^{i (\bbox{k}_{\|} + \bbox{G}_{\|})\cdot
\bbox{r}}\; \Psi_{\mu}(z)
\]
with corresponding energy
\[
E(\bbox{G}_{\|}) = E_{\mu} + 
\frac{ \hbar^2 (\bbox{k}_{\|} + \bbox{G}_{\|})^2}{2m}.
\] 
A surface resonance would
correspond to a $\Psi_{\mu}(z)$ which is localized around the surface. \\
Let us now consider the subspace of eigenstates
of $V_{av}(z)$  which comprises\\
a) the (discrete) surface resonance state $\Psi_{\bbox{G}_{\|},\mu}$
and \\
b) the continuum of states $\Psi(\bbox{k}_{\|},k_\perp)$
which evolve into plane waves with momentum
$(\bbox{k}_{\|}, k_\perp)$ asymptotically far away
from the surface. 
The periodic surface potential (i.e. the term $V_1$)
then provides a mechanism for mixing $\Psi_{\bbox{G}_{\|},\mu}$
and the continuum. Moreover,
dipole transitions of an electron from the valence band 
into both a continuum state and into the surface resonance
(whereby the latter alternative is possibly even more probable
due to the special radiation characteristics of a ZRS) are possible.
We thus have all ingredients for the standard Fano-type
resonance (see Figure \ref{fano})
and we would thus expect a pronounced peak
\noindent
in an intensity-versus-photon-energy curve whenever the kinetic energy
of the photoelectrons approximately matches the energy of a
surface resonance. Neglecting the (presumably small)
energy $E_{\mu}$ this will happen when the kinetic energy $T$
obeys the emergence condition (\ref{emergence}) for some $\bbox{G}_{\|}$.\\
A strong oscillation of the intensity
of the ZRS-derived first ionization states at $(\pi/2,\pi/2)$
and $(0.7\pi,0)$ in Sr$_2$CuCl$_2$O$_2$ as a function of photon energy 
has indeed been found recently by
D\"urr {\em at al.}\cite{duerr}.
Their data show a total of four maxima of intensity
in the photon-energy range $10-70\;eV$.
Much unlike the absorption cross section oscillations seen in EXAFS, these
maxima are separated by near-zeroes of the
intensity for certain photon energies.\\
One interpretation which might come to mind
would be interference between the directly emitted electron wave
and partial waves which have been reflected from 
neighboring atoms, similar as the variations in absorption cross
section seen in EXAFS.
If we neglect the energy dependence of the scattering phase shift at
the hypothetical scatterers,
the difference $d$ in pathlength between the interfering
partial waves would then obey
\[
d(k^{max}_{\nu+1} - k^{max}_{\nu})=2\pi
\]
where $k^{max}_{\nu}$ is the free electron wavevector at the
$\nu^{th}$ maximum of intensity.
From the four maxima observed by D\"urr {\em at al.}
at  $(0.7\pi,0)$, one obtains three estimates for $d$:
$10.6$, $12.6$, $9.5$ $\AA$. This is in any case much longer
than the Cu-O bond length of $2$ $\AA$. The only
`natural length' in Sr$_2$CuCl$_2$O$_2$ which
would give a comparable distance is the
distance between the two inequivalent CuO$_2$ planes, which is
$7$ $\AA$. One might therefore be tempted to explain the oscillation
as being due to electrons from the topmost CuO$_2$ plane
being reflected at the first  CuO$_2$ plane below. This would
give a difference in pathway of approximately $14$ $\AA$, the
discrepancy with the values given above could possibly
be explained by an energy
dependence of the phase shift upon reflection.
On the other hand it is quite obvious that the reflected wave,
having to travel a total of $14$ $\AA$ through the interior of
the solid and having undergone one reflection, would have
a considerably smaller amplitude than the primary wave. 
In this picture
it is then hard to explain why the intensity at the minima
is so close to zero. We therefore believe that interference is
the correct explanation of the strong
oscillations.\\
Let us assume on the other hand, that the maxima are due to
Fano-resonances of the type discussed above.
As already mentioned above, we would expect a surface resonance state to 
form whenever
the emergence condition (\ref{emergence}) is fulfilled approximately
for some reciprocal lattice vector $\bbox{G}_{\|}$.
\begin{figure}
  \begin{center}
    \epsfig{file=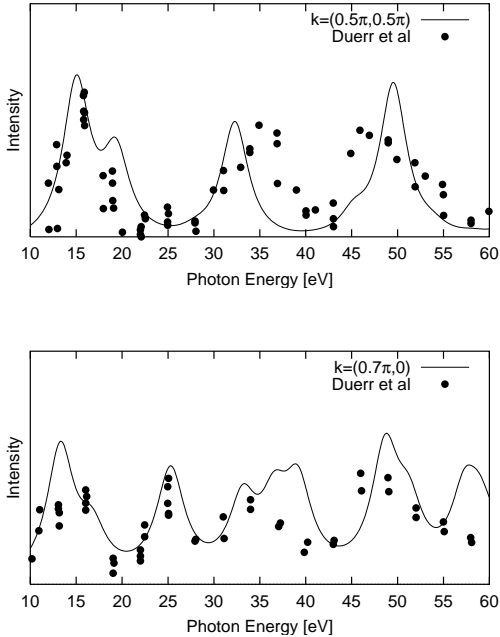,width=7.0cm}    
  \end{center}
\narrowtext
\caption[]{Intensities of the first ionization state at
$(\frac{\pi}{2},\frac{\pi}{2})$ (top) and $(0.7\pi,0)$ (bottom)
in  Sr$_2$CuCl$_2$O$_2$ as a function of photon energy. 
The dots are experimental data from Ref. \cite{duerr}, the
line is calculated using (\ref{theoint}).}
\label{duerrcomp}
\end{figure}
\noindent
As a rough approximation, we moreover assume that the
probability for the dipole transition from the
CuO$_2$ plane into the surface resonance state
is proportional the `transition matrix element'
$m_{ZRS}$ into a plane wave state with momentum 
$(\bbox{k}_{\|}+\bbox{G}_{\|}, k_\perp=0)$.
Neglecting the form of the Fano lineshape (which would be hard to compute
anyway because we do not know the Fourier coefficients of
the potential $V_1$) we then expect an energy dependence of
the intensity to be roughly given by
\begin{equation}
I(h\nu) \approx \sum_{\bbox{G}_{\|}} |m_{ZRS}(\bbox{k}_{\|}+\bbox{G}_{\|})|^2
\delta(T -  \frac{\hbar^2}{2m}(\bbox{k}_{\|} + \bbox{G}_{\|})^2).
\label{theoint}
\end{equation}
Actually injection into the surface state should be possible
if the kinetic energy is in a narrow window above the 
emergence condition, whence for simplicity we replace the
$\delta$-functions by Lorentzians.
Figure \ref{duerrcomp} then compares this simple estimate
to the data of 
D\"urr {\em at al.}\cite{duerr}. 
Thereby we have again assumed that $T=h\nu - 8\;eV$.
The agreement with experiment is satisfactory,
given the rather crude nature of our estimate for the intensity.
Given the fact that surface resonances in Sr$_2$CuCl$_2$O$_2$
have been established conclusively by the EELS-work of
Pothuizen\cite{hans_thesis}
we believe that the present
interpretation of the intensity variations is the most plausible one.
The question to whether such surface resonance states exist also in
metallic Bi2212 must be clarified by experiment.
\section{Apparent Fermi surfaces}
\begin{figure}
  \begin{center}
    \epsfig{file=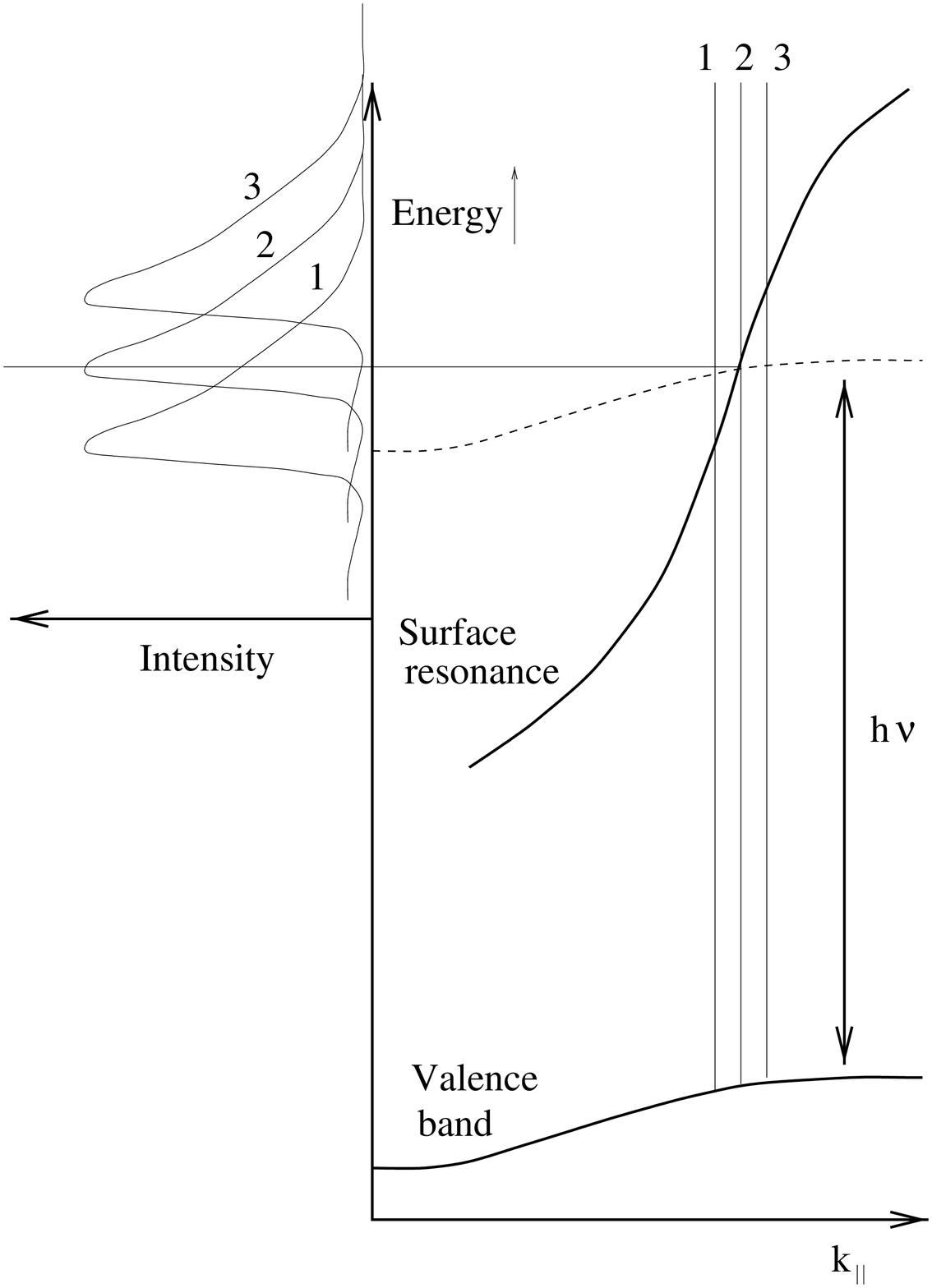,width=6.0cm,angle=0.0}
  \end{center}
\narrowtext
\caption{Creation of an `apparent Fermi surface' by
a combination of Fano resonance and surface state dispersion.}
\label{pseudo}
\end{figure}

\begin{figure}
  \begin{center}
    \epsfig{file=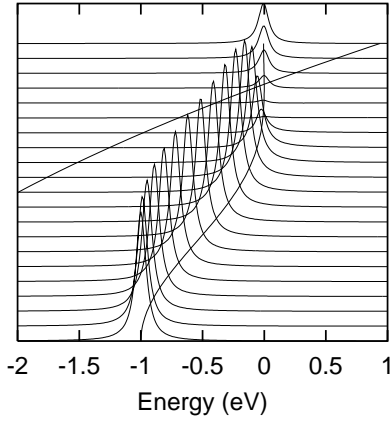,width=6.0cm,angle=0.0}    
  \end{center}
\narrowtext
\caption{Spectral intensity computed from (\ref{arpesspec}.)
The spectra are calculated for equidistant
momenta along a line through the
origin $(0,0)$ (lowermost spectrum) which forms an angle of $10^o$ 
with the $(1,0)$ direction.
The quasiparticle dispersion $\epsilon(\bbox{k}_{\|})$ and the dispersion
of the surface resonance, shifted downwards by the photon energy
$h\nu$, are given by lines.}
\label{fanodisp}
\end{figure}

\begin{figure}
  \begin{center}
    \epsfig{file=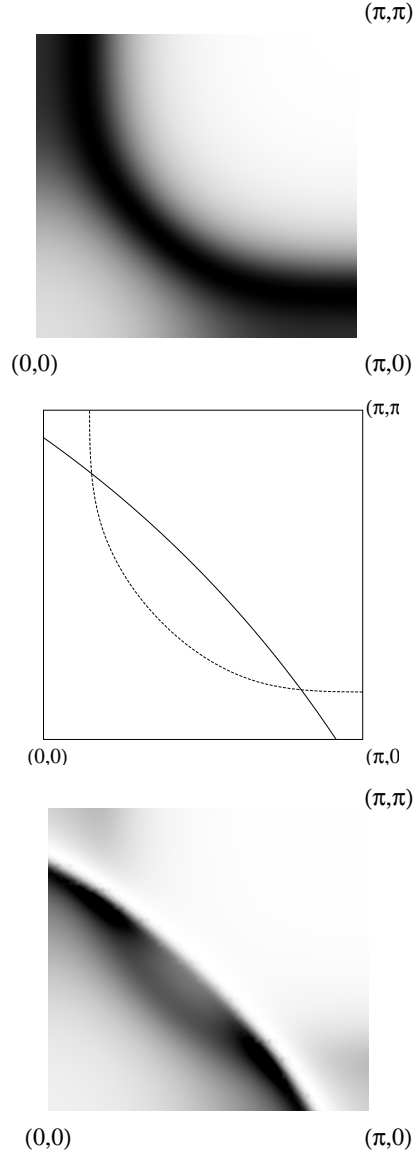,width=6.0cm,angle=0.0}
  \end{center}
\narrowtext
\caption{(a) Gray scale plot of the spectral intensity
immediately below $E_F$, computed from (\ref{arpesspec})
with $A(\omega)=1$.\\
(b) Fermi surface (dashed line) and constant energy contour of the surface state
(full line).\\
(c) Spectral intensity in the presence of the surface state, i.e.
(\ref{arpesspec}) with the true $A(\omega)$.}
\label{denplo}
\end{figure}

\begin{figure}
  \begin{center}
    \epsfig{file=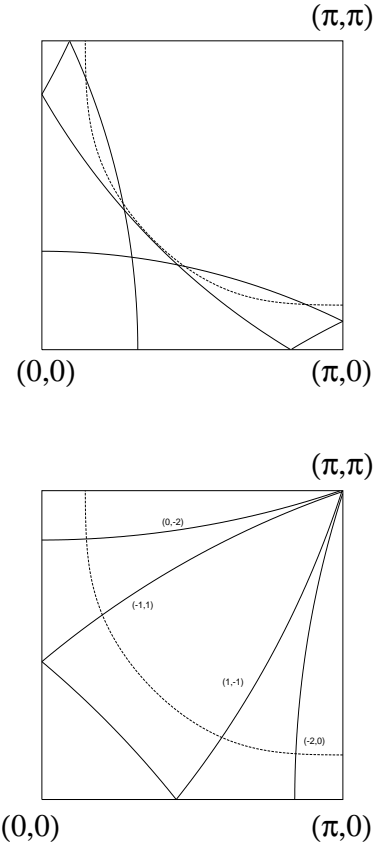,width=5.0cm,angle=0.0}
  \end{center}
\narrowtext
\caption{(a) The full lines give the
constant energy contours for a free electron-like
surface resonance state for kinetic energy $22\;eV-8\;eV$ (a)
and $34\;eV-8\;eV$ (b). The dashed line is the `Fermi surface'
seen in Bi2212.}
\label{circles}
\end{figure}

Let us now consider what happens in the above picture
if we vary the in-plane momentum $\bbox{k}_{\|}$.
It is important to notice from the outset, that the group velocity of the
surface state, ${\bf v}_{\|}= \frac{\hbar^2}{m}(\bbox{k}_{\|}+
\bbox{G}_{\|})$, 
is much larger than that of the valence band.
Thus, if we shift the valence band `upwards' by the 
photon energy $h\nu$, this replica and the surface state my intersect the
surface resonance dispersion at some momentum (see Figure \ref{pseudo}). 
Note that here the binding energy of the O2p level must be incorporated into
the valence band energy.
For each $\bbox{k}_{\|}$ we now assume that the Fano-type interference
between the surface resonance state and the
continuum occurs. The maximum of the resonance curve thereby will
roughly follow the dispersion of the surface state.
In an ARPES experiment we then obviously probe the intensity
along the Fano-curve at an energy which corresponds to
the shifted valence band (see Figure \ref{pseudo}). 
Obviously this leads to a drastic variation
of the observed photoelectron intensity:
at the $\bbox{k}_{\|}$ labeled 1, the point where the resonance
curve is probed is on the ascending side of the resonance, whence we
observe a moderately large 
\noindent
intensity. At the second momentum, we are right at the maximum of the
respective Fano-resonance, so the intensity will be high.
At the momentum labeled 3, however, we are already at the minimum of the
Fano-resonance, where the intensity is approximately zero.
We thus would observe a dramatic drop
in the observed intensity over a relatively small distance $\Delta k$
in the $\bbox{k}_{\|}$-plane.
More precisely, the sharpness of the drop is determined by the
width $W$ of the Fano-curve and the difference $\Delta v_g$
in group velocity between the valence band and the surface resonance state:
$\Delta k\approx W/\Delta v_g$.\\
To be more quantitative, we present a simple model
calculation of the ARPES spectra to be expected in the presence of
a free-electron-like surface resonance.
We label the surface resonance state
as $|0\rangle$ and the continuum of LEED states by their `perpendicular'
momentum: $|k_\perp\rangle$. The Hamiltonian then reads
\begin{eqnarray}
H_0 &=& |0\rangle E(\bbox{k}_{\|},\bbox{G}_{\|}) \langle 0|
+ \sum_{k_\perp} |k_\perp\rangle  \frac{\hbar^2 k_\perp^2}{2m}
\langle k_\perp |\\
H_1 &=&  \sum_{k_\perp} (  |k_\perp\rangle \frac{V}{\sqrt{L}}
 \langle 0| + H.c.),\nonumber \\
 E(\bbox{k}_{\|},\bbox{G}_{\|}) &=&  \frac{ \hbar^2 (\bbox{k}_{\|} + \bbox{G}_{\|})^2}{2m}.
\end{eqnarray}
We assume that the vacuum consists of a volume spanned by $N\times N$ planar
unit cells and thickness $L$ perpendicular to the surface.
We write the mixing matrix element as $ V/\sqrt{L}$ so as to explicitly
isolate the scaling of the matrix elements
with $L$. $V$ should be independent of $L$ and since we assume it to be 
independent of $k_\perp$ we may also assume it to be real.\\
We then obtain the resolvent operator
\begin{eqnarray*}
R_{0,0}(\omega) &=& \langle 0| (\omega -i0^+ - H)^{-1} |0\rangle \nonumber \\
&=& \frac{1}{\hbar \tilde{\omega} - E_0(\bbox{k}_{\|}) -
\Sigma(\tilde{\omega})} \nonumber \\
\hbar\tilde{\omega} &=& \hbar \omega - \frac{\hbar^2}{2m}\bbox{k}_{\|}^2 \\ 
\Sigma(\omega) &=& i \frac{V^2}{4\pi^2}
\sqrt{\frac{2m}{\hbar^3 \omega}} \nonumber \\
&=& i\;v_0 \sqrt{ \frac{\omega_0}{\omega}} 
\end{eqnarray*}
Here $v_0$ and $\omega_0$ have the dimension of energy and frequency,
respectively; only one of them can be chosen independently, we choose
$\omega_0 = 1eV/\hbar$ whence $v_0$ is a measure for the strength
of the mixing between surface resonance and continuum.\\
A possible final state obtained by dipole transition
of an electron from the valence band then is
\begin{equation}
|\Phi\rangle = T_0\; |0\rangle + \frac{T_1}{\sqrt{L}}\; \sum_{k_\perp} |k_\perp\rangle
\end{equation}
where we have again explicitly isolated the scaling of the transition matrix 
elements with $L$.
Both matrix elements are assumed to be real.
The measured intensity then becomes
\begin{eqnarray}
A(\omega) &=& \frac{1}{\pi} \Im \;R(\omega)\nonumber \\
R(\omega) &=&  \langle \Phi| (\omega -i0^+ - H)^{-1} |\Phi\rangle
\nonumber \\
&=& 
\frac{ \left(T_0 + \tilde{T}_1\Sigma(\tilde{\omega})\right)^2}
{\hbar \tilde{\omega} - 
E(\bbox{k}_{\|},\bbox{G}_{\|}) -
\Sigma(\tilde{\omega})} + \tilde{T}_1^2 \Sigma(\tilde{\omega})
\end{eqnarray}
where we have introduced $\tilde{T}_1= T_1/V$.
Taking into account the dispersion of the initial state
as well as its finite lifetime (which we describe by a Lorentzian
broadening $\Gamma$) we approximate the measured spectrum for some momentum
$\bbox{k}_{\|}$ as
\begin{equation}
I(\omega) = \frac{\Gamma}{(\omega- \epsilon(\bbox{k}_{\|}))^2 + \Gamma^2}\cdot
A(\omega+h\nu)
\label{arpesspec}
\end{equation}
A simulated ARPES spectrum is then shown in Figure \ref{fanodisp}.
For simplicity we have chosen $\epsilon(\bbox{k}_{\|})$ to be
the simple SDW-like dispersion 
\begin{eqnarray}
\epsilon(\bbox{k}_{\|}) &=& -t_{eff} (\cos{k_x} + \cos(k_y))^2 
\end{eqnarray}
with $t_{eff}=0.25\;eV$.
It has been assumed that this band is completely filled,
that means we would not have any true Fermi surface at all.
Despite this, the ARPES spectra, which is simulated
by using (\ref{arpesspec})
shows a sharp drop in intensity at the intersection of
the surface state dispersion $E(\bbox{k}_{\|},\bbox{G}_{\|})$
shifted downward by the photon energy. 
Here $\bbox{G}_{\|}=(0,2\pi)$.
\noindent
The photon energy in the above example has been chosen deliberately
such that the surface state dispersion cuts through the quasiparticle 
dispersion near the top of the latter, so as to produce the
impression of a Fermi surface.
Usually this would happen only by coincidence for very few
photon energies.
At this point is has to be remembered, however, that the dispersion of
the valence band in cuprate superconductors  seems
to show an extended region with practically no dispersion in the region around
$(\pi,0)$. This band portion moreover is energetically
immediately below the chemical potential.
The almost complete lack of dispersion in this region then makes it
necessary to rely mostly on `intensity drops' of the valence band
in assigning a Fermi surface, and it is quite obvious that whenever 
the surface state dispersion, shifted downward by the photon energy,
happens to cut through the flat-band region 
for the photon energy in question, the resulting drop
in intensity will be almost indistinguishable from a true Fermi surface.
To illustrate this effect we have calculated the intensity
within a window of $10\;meV$ below $E_F$ for the `standard' 
dispersion with an extended van-Hove singularity\cite{Radtke}
\begin{eqnarray}
\epsilon_{QP}(\bbox{k}_{\|}) &=& c 
-\frac{t}{2} \left( \cos(k_x) + \cos(k_y) \right)
\nonumber \\
&+& t' \cos(k_x)\cos(k_y) \nonumber \\
&+& \frac{t''}{2} \left( \cos(2k_x) + \cos(2k_y) \right)
\end{eqnarray}
which (for $t=0.5\;eV$, $t'=0.15\;eV$ and $t''=-0.05\;eV$)
produces the well-known `$22\;eV$-Fermi surface', together
with the `flat bands' around $(\pi,0)$. The constant $c$
incorporates a constant shift due to
the binding energy of the O 2p orbitals.
Figures \ref{denplo} then shows the integrated intensity in a window of
$10\;meV$ below the Fermi energy - a plot which is by now a standard
method to discuss Fermi surfaces. In Figure \ref{denplo}a
the Fano-curve $A(\omega)$ in (\ref{arpesspec}) is replaced by unity,
and we see the expected spectral weight map in which the Fermi surface can
be clearly identified. 
\noindent
The latter is shown in Figure \ref{denplo}b.
In Figure \ref{denplo}c, on the other hand,
we have included the Fano curve $A(\omega)$ for the surface state with 
$\bbox{G}_{\|}=(-2\pi,-2\pi)$ (assuming a free 2D-electron dispersion
with lattice constant $2.82 Angstrom$) and a kinetic energy of
$T=33eV$. The corresponding constant energy-contour
$\frac{4\pi^2\hbar^2}{2ma^2}( \bbox{k}+\bbox{G})^2 =33eV$
is also shown in Figure \ref{circles}. One can see quite clearly then,
that the contour attenuates the true Fermi surface on its `backside'
and strongly enhances the intensity at its `frontside' thus
creating the rather perfect impression of a Fermi surface arc
which intersects the $(1,0)$ direction at approximately
$(0.8\pi,0)$. On the other hand in our model
the true Fermi surface is the one seen in Figure \ref{denplo}a.
The combination of Fano resonance and flat band portion near
$(\pi,0)$ thus produces an `apparent Fermi surface'
which is entirely artificial. \\
To compare with experimental data in more detail, we recall that
upon neglecting the dispersion of the valence band state as compared
to that of the surface resonance, the momenta in the  $\bbox{k}_{\|}$-plane
where intersections as the one in Figure
\ref{pseudo} occur have to obey the equation
\[
T = \frac{ \hbar^2 (\bbox{k}_{\|} + \bbox{G})^2}{2m},
\]
where $T$ denotes the kinetic energy of the ejected electron.
In other words: the resonant enhancement of the
ARPES intensity occurs along a constant
energy contour of the surface resonance dispersion.
If we stick to our free-electron approximation, these
are simply circles in the $\bbox{k}_{\|}$-plane centered
on $-\bbox{G}$. Figure \ref{circles} then shows these
contours for two different photon energies, namely $22\;eV$ 
and $34\;ev$. Thereby we have again assumed that $T=h\nu-8\;eV$.
Also shown is the standard `Fermi surface', determined at
$22\;eV$.
It is then obvious that the `$22\;eV$-Fermi surface'
always is close to some portion of a surface resonance contour.
This would imply that for this photon energy
most potions of the Fermi surface could be enhanced -
it might also be taken to suggest, however, that some 
Fermi surface portions near $(\pi,0)$
are actually artificial. 
It is hard to give a more detailed discussion, because even slight
deviations from the free electron dispersion for the surface resonance
will have a major impact on the energy contours.
For $34\;eV$, on the other hand, the Fermi surface obviously is `cut off'
near $(\pi,0)$ by the contour for ${\bf G}=2\pi\;(-2,0)$.
The Fermi surface seen at $34\;eV$ then might be obtained by following the
true Fermi contour (dashed line) until the intersection with the
$(-2,0)$-contour, and then following the latter (where the
ARPES intensity is high due to the resonance effect and
the fact there are states immediately below $\mu$ due to the flat band
around $(\pi,0)$). 
In this way one would obtain a `Fermi contour' which is quite similar
to the one actually observed at $34\;eV$.
For even higher photon energy there are too many
$\bbox{G}$'s which contribute their own circles, 
so that one cannot give a meaningful discussion.
\noindent
The question which of the different
Fermi surface portions observed in Bi2212 are artificial and which 
are real clearly needs a detailed experimental study. 
A very simple way would be to vary the photon energy in small steps of
(e.g.) $1eV$, in which case one should observe a continuous
drift of the `Fermi surface' for those parts which are
artificial.
Unfortunately the present understanding of
surface resonances in cuprate materials is much too rudimentary to make
any theoretical prediction.
Not even knowing any details of their dispersion, we have to content 
ourselves with the very much oversimplified model calculations
outlined above to show what might happen.
Even these simple model calculations do show rather
clearly, however, that simply plotting intensity maps near $E_F$
or identifying `Fermi surface crossings' near $(\pi,0)$ by
drops of spectral weight is not really an adequate means to
clarify this issue. 

\section{Conclusion}
In summary, we have presented a discussion of ARPES
intensities from the CuO$_2$ plane. Thereby we could actually
address only a few (but hopefully the most important)
aspects of the problem. \\
To begin with, the 
geometry of the orbitals, which form the first ionization
states of the CuO$_2$ plane, leads to a pronounced anisotropy of the
photoelectron current. This makes itself felt in a strong
dependence of the intensity on the
polarization of the incoming light. This may be exploited to
enhance the intensity seen in an ARPES experiment
by choosing appropriate photon polarization or going to a higher
Brillouin zone. We have presented a simple model to guess `optimized'
values of the polarization and the Brillouin zone to
obtain maximum intensity for a given ${\bf k}$-point.\\
Moreover there is a strong
preference for a Zhang-Rice singlet to emit photoelectrons 
at small angles relative to the plane.
This in turn may lead to a injection of the photoelectrons into
states located at the surface of the sample, so-called surface resonances.
Such surface resonances are well-established in at least one
cuprate-related material, namely Sr$_2$CuCl$_2$O$_2$.
We have proposed to explain the pronounced photon-energy dependence of
the ARPES intensities in this material by these surface resonances.
Next, we have presented a simple model calculation which shows how
such surface resonances may lead to apparent Fermi surfaces in the
flat-band region around $(\pi,0)$. This may be one
explanation for the apparently different Fermi surface topology
seen in Bi2212 at $22\;eV$ and at $34\;eV$, possibly
also in an interplay with bilayer-splitting\cite{Armitage,Chuang}.
In any case the existence or non-existence of surface-resonances 
should be clarified experimentally, since the identification of any
major Fermi surface portion near $(\pi,0)$ as being `artificial'
amy have major implications for our understanding of cuprate materials.\\
Acknowledgment: Instructive discussions with V. Borisenko, J. Fink, M. Golden
and A. Fleszar are most gratefully acknowledged.
This work is supported by the projects DFG HA1537/20-2,
KONWIHR OOPCV and BMBF 05SB8WWA1.
Support by computational facilities of HLRS Stuttgart and LRZ Munich
is acknowledged.
 
\section{Appendix I:Calculation of Radial Matrix Elements}
The actual calculation of the radial matrix elements $R_{l,l'}$
prove to be one of the most crucial points when making contact
with experimental results. Functional forms of effective potentials
for the CuO$_2$ compound are not known. A simple approach with a
hydrogen-like potential seems doubtful, if one attempts cover at
least a certain amount of physical reality of the photoemission 
process. 
A good starting point for an at 
least qualitative derivation of the matrix elements are the radial 
wave functions that can be obtained from density functional calculations, 
which include information about the electron-electron exchange 
correlation and the screening of the nuclear charge. 
The radial matrix elements used in this paper are calculated
in the local-density approximation, 
where one solves the effective one electron Schr\"odinger equation
\begin{equation}
  \label{eq:sic1}
\left(\frac{1}{2} \nabla^2 + v^{\alpha,\sigma}_{\mbox{eff}} \left( {\bf r} \right) \right) \psi_{\alpha,\sigma}
=\epsilon^{LDA}_{\alpha,\sigma}\psi_{\alpha,\sigma}\left( {\bf r}\right).  
\end{equation}
The potential $v^{\alpha,\sigma}_{\mbox{eff}} \left( {\bf r}\right)$ 
is a functional of the electron density, given by
\[
v^{\alpha,\sigma}_{\mbox{eff}} \left( {\bf r}\right)
=\left( 
  v\left( {\bf r}\right) + u\left([n]; {\bf r}\right)
  +v^{\sigma}_{\mbox{xc}} \left(\left[ n_\uparrow,n_\downarrow \right];{\bf r} \right)
\right).
\]
$v({\bf r})$ is the full nuclear potential and $u\left([n]; {\bf r}\right)$ the 
direct Coulomb-potential $\int d{\bf r}' n\left({\bf r}' \right)/\left| {\bf r}-{\bf r}' \right|$.
The exchange-correlation functional  
$v^{\sigma}_{\mbox{xc}} \left(\left[ n_\uparrow,n_\downarrow \right];{\bf r} \right)$
is a parameterized after Ref. \cite{perdew}.
Equation (\ref{eq:sic1}) is solved selfconsistently to convergence of the total energy, 
resulting in a set of energy eigenvalues $\epsilon_n$ and radial wave functions $R_{n,l}(r)$, 
which we used to calculate
\[
R_{ll'}(E)=\int dr  r^3 R_{E,l}(r) R_{n,l}(r).
\]
Since we assume emitted electrons to be approximated by plane waves 
at the detection point, additional information on the outgoing wave function is needed
in form of the phase shift compared to the asymptotic behaviour of 
\[
R^{asym}_{El}(r) \approx 2 \frac{\sin\left(kr-\frac{l\pi}{2}+\delta_l \right)}{r}, 
\qquad r \rightarrow \infty,
\]
defining the interference of wave functions emitted by Op$_{x,y}$ and Cud$_{x^2-y^2}$ orbitals.
The phase shift can easily be found by comparing the logarithmic derivatives
\[
\left(\frac{\partial}{\partial r}R\left( r \right) \right)/R\left( r \right)
=\left(\frac{\partial}{\partial r}R^{asym}_{El}\left( r \right) \right)/R^{asym}_{El}\left( r \right)
\]
at a sufficiently large radius where 
\[
\frac{\partial}{\partial r} v^{\alpha,\sigma}_{\mbox{eff}} \left( {\bf r}  \right) \approx 0.
\]

\begin{figure}[]
  \begin{center}
    \epsfig{file=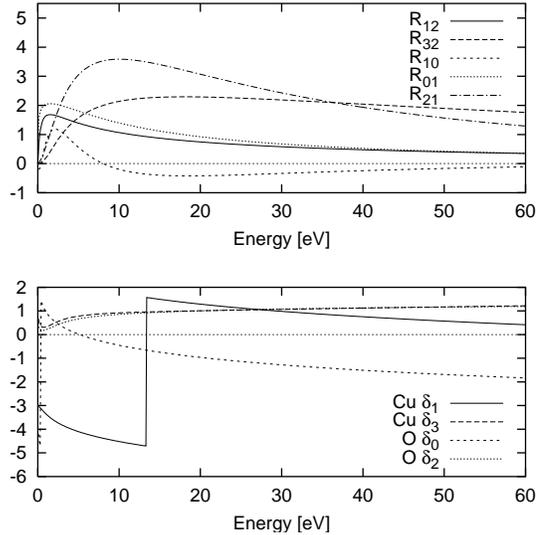}
  \end{center}
\narrowtext
\caption{Radial integrals 
  R$_{l,l'}$ (top) and Scattering phases $\delta_{l,l'}$ (bottom)  
  as calculated from density functional theory.}
\label{ri}
\end{figure}
\end{multicols}
\widetext
\section{Appendix II}
Here we present in more detail the calculation of the vectors $\bbox{v}$ of
interest to us.
For the CEF initial states of interest to us,
such as $p_x$, $p_y$ and $d_{x^2-y^2}$,
the coefficients $c_{\alpha m'}$ take the form
\[
c_{\alpha m'}= \frac{\zeta}{\sqrt{2}}\;
( \delta_{m',\nu} \pm \delta_{m',-\nu}),
\]
(see Table I) whence we find that
\begin{eqnarray}
\sum_{m}\; d_{lm}\; Y_{lm}(\bbox{k}^0) 
&=& \sqrt{\frac{4\pi}{3}}
\sum_{\mu=-1}^1\;  Y_{1,\mu}^*(\bbox{\epsilon})\; A_{l,\mu}(\bbox{k}^0)
\nonumber \\
 A_{l,\mu}(\bbox{k}^0) &=& 
\frac{\zeta}{\sqrt{2}} \;
\left( Y_{l,\mu+\nu}(\bbox{k}^0)\; c^1(l, \mu+\nu; l', \nu) \pm
Y_{l,\mu-\nu}(\bbox{k}^0)\; c^1(l, \mu-\nu; l',-\nu) \right)
\label{prod}
\end{eqnarray}
We now rewrite the scalar product
\[
 \sqrt{\frac{4\pi}{3}}\sum_{\mu=-1}^1\;  Y_{1,\mu}^*(\bbox{\epsilon})\; A_\mu^l
= \bbox{ \epsilon} \cdot \bbox{A}_l
\]
and by use of the property $c^1(l,-m; l',-m') = c^1(l,m; l',m')$ we obtain
the components of $\bbox{A}^l$:
\begin{eqnarray}
A_{l,x} &=& \frac{1}{\sqrt{2}}
\;[\; c^1(l,\nu-1;l',\nu)\; \frac{\zeta}{\sqrt{2}}\;
(Y_{l,\nu-1} \mp Y_{l,-(\nu-1)}) 
-   c^1(l,\nu+1;l',\nu)\; \frac{\zeta}{\sqrt{2}}\;
(Y_{l,\nu+1} \mp Y_{l,-(\nu+1)}) \; ],
\nonumber \\
A_{l,y} &=& \frac{i}{\sqrt{2}}
\; [\; c^1(l,\nu-1;l',\nu)\; \frac{\zeta}{\sqrt{2}}\;
(Y_{l,\nu-1} \pm Y_{l,-(\nu-1)}) 
+   c^1(l,\nu+1;l',\nu)\; \frac{\zeta}{\sqrt{2}}\;
(Y_{l,\nu+1} \pm Y_{l,-(\nu+1)}) \; ], 
\nonumber \\
A_{l,z} &=&  c^1(l\nu;l'\nu)\; \frac{\zeta}{\sqrt{2}}
(Y_{l,\nu} \pm Y_{l,-\nu}).
\end{eqnarray}
Defining $\tilde{R}_{l,l'}(E) = \;e^{i\delta_{l}}  R_{l,l'}(E)$
we now readily
arrive at the following expressions for photocurrent:
\begin{eqnarray}
{\bf j} &=&  \frac{4N \hbar {\bf k}}{m}\; |\bbox{v} \cdot \bbox{\epsilon} |^2,
\nonumber \\
\bbox{v} &=&  \sum_{l=l'\pm1} (-i)^{l+1}\;\tilde{R}_{l,l'}(E)\;\bbox{A}_l.
\end{eqnarray}
Here the vector $\bbox{v}$ depends on the type
of initial orbital. Straightforward algebra then yields the following 
expressions for the vectors  $\bbox{v}$:
\begin{eqnarray}
\bbox{v}(p_x) &=& 
 \sqrt{ \frac{1}{15} } \tilde{R}_{21} \left(
\begin{array}{c}
\sqrt{\frac{5}{4\pi}} \frac{\tilde{R}_{01}}{\tilde{R}_{21}} + 
d_{3z^2-r^2}(\bbox{k}^0) - \sqrt{3}  d_{x^2-y^2}(\bbox{k}^0) \\
-\sqrt{3} d_{xy}(\bbox{k}^0) \\
-\sqrt{3} d_{xz}(\bbox{k}^0)
\end{array} \right)
\nonumber \\
\bbox{v}(p_y) &=&
\sqrt{ \frac{1}{15} } \tilde{R}_{21} \left(
\begin{array}{c}
-\sqrt{3} d_{xy}(\bbox{k}^0) \\
\sqrt{\frac{5}{4\pi}}\frac{\tilde{R}_{01}}{\tilde{R}_{21}}  + 
d_{3z^2-r^2}(\bbox{k}^0) + \sqrt{3}  d_{x^2-y^2}(\bbox{k}^0) \\
-\sqrt{3} d_{yz}(\bbox{k}^0)
\end{array} \right)
\nonumber \\
\bbox{v}(p_z) &=&
\sqrt{ \frac{1}{15} } \tilde{R}_{21} \left(
\begin{array}{c}
-\sqrt{3} d_{xz}(\bbox{k}^0) \\
-\sqrt{3} d_{yz}(\bbox{k}^0) \\
\sqrt{\frac{5}{4\pi}}\frac{\tilde{R}_{01}}{\tilde{R}_{21}} +
\sqrt{4} d_{3z^2-r^2}(\bbox{k}^0)
\end{array} \right)
\end{eqnarray}
With the exception of the numerical prefactors,
most of the above formulas could have been guessed
from general principles:
there are only transitions into $s$-like and $d$-like
partial waves (i.e. the dipole selection rule $\Delta L = \pm 1$!)
and acting e.g. with an electric field in $x$ direction onto a
$p_x$ -orbital produces only $s$-like, $d_{x^2-y^2}$-like and
$d_{3z^2-r^2}$-like partial waves (which have even parity under
reflection by the $z-y$-plane), whereas acting with
a field in $y$ direction on $p_x$ can only
produce the $d_{xy}$ partial wave. 
We therefore believe that
much of the formula remains true even if more realistic wave functions
for the final states are chosen.\\
Similarly, we find for the matrix element of the $d$-like orbitals:
\begin{eqnarray}
  \label{eq:d0}
  {\bf v}\left(d_{3z^2-r^2} \right)&=& i\left( \matrix{ 
      \frac{1}{\sqrt{30}} \tilde{R}_{12} p_x
      +   \sqrt{\frac{3}{35}}\tilde{R}_{32} f_{5x(z^2-r^2)}
      \cr 
      \frac{1}{\sqrt{30}} \tilde{R}_{12} p_y
      + \sqrt{\frac{3}{35}}\tilde{R}_{32} f_{5y(z^2-r^2)} 
      \cr
      -\frac{2}{\sqrt{15}} \tilde{R}_{12} p_z 
      +\frac{3}{\sqrt{35}}\tilde{R}_{32} f_{z(5z^2-3r^2)}      }\right)
\end{eqnarray}

\begin{eqnarray}
  \label{eq:dyz}
  {\bf v}\left(d_{yz} \right)&=& i
\left( \matrix{
      \frac{1}{\sqrt{7}} \tilde{R}_{32} f_{xyz} \cr
      -\frac{1}{\sqrt{5}} \tilde{R}_{12} p_z - \sqrt{\frac{3}{35}} \tilde{R}_{32} f_{5z^3-3z} -\frac{1}{\sqrt{7}} \tilde{R}_{32} f_{z(x^2-y^2)} \cr
      -\frac{1}{\sqrt{5}} \tilde{R}_{12} p_y +\sqrt{\frac{8}{35}} \tilde{R}_{32} f_{y(5z^2-1)}} \right)
\end{eqnarray}

\begin{eqnarray}
  \label{eq:dxz}
  {\bf v}\left(d_{xz} \right)&=&  i   
  \left( \matrix{
      -\frac{1}{\sqrt{5}} \tilde{R}_{12} p_z-\sqrt{\frac{3}{35}} \tilde{R}_{32} f_{5z^2-3z} +\frac{1}{\sqrt{7}} \tilde{R}_{32} f_{z(x^2-y^2)} \cr
      \frac{1}{\sqrt{7}} \tilde{R}_{32} f_{xyz} \cr
      -\frac{1}{\sqrt{5}} \tilde{R}_{12} p_x +\sqrt{\frac{8}{35}} \tilde{R}_{32} f_{x(5z^2-1)}
      } \right)
\end{eqnarray}
\begin{eqnarray}
  \label{eq:dx2-y2}
  {\bf v}\left(d_{x^2-y^2} \right)&=&i
  \left( \matrix{
      -\frac{1}{\sqrt{5}} \tilde{R}_{12} p_x -\frac{1}{\sqrt{70}} \tilde{R}_{32} f_{x(5z^2-1)} 
      +\sqrt{\frac{3}{14}} \tilde{R}_{32} f_{x^3-3xy^2} \cr
      \frac{1}{\sqrt{5}} \tilde{R}_{12} p_y +\frac{1}{\sqrt{70}} \tilde{R}_{32} f_{y(5z^2-1)} 
      +\sqrt{\frac{3}{14}} \tilde{R}_{32} f_{3yx^2-y^3} \cr
      \frac{1}{\sqrt{7}} \tilde{R}_{32} f_{z(x^2-y^2)}
      } \right)
\end{eqnarray}

\begin{eqnarray}
  \label{eq:dxy}
  {\bf v}\left(d_{xy} \right)&=&i
  \left( \matrix{
      -\frac{1}{\sqrt{5}} \tilde{R}_{12} p_y -\frac{1}{\sqrt{70}} \tilde{R}_{32} f_{y(5z^2-1)} 
      +\sqrt{\frac{3}{14}} \tilde{R}_{32} f_{yx^2-y^3} \cr
      -\frac{1}{\sqrt{5}} \tilde{R}_{12} p_x -\frac{1}{\sqrt{70}} \tilde{R}_{32} f_{x(5z^2-1)} 
      -\sqrt{\frac{3}{14}} \tilde{R}_{32} f_{x^3-xy^2} \cr
      \frac{1}{\sqrt{7}} \tilde{R}_{32} f_{xyz}
      } \right)
\end{eqnarray}
With the exception of the numerical prefactors,
most of the above formulas could have been guessed
from general principles:
there are only transitions into $s$-like and $d$-like
partial waves (i.e. the dipole selection rule $\Delta L = \pm 1$!)
and acting e.g. with an electric field in $x$ direction onto a
$p_x$ -orbital produces only $s$-like, $d_{x^2-y^2}$-like and
$d_{3z^2-r^2}$-like partial waves (which have even parity under
reflection by the $z-y$-plane), whereas acting with
a field in $y$ direction on $p_x$ can only
produce the $d_{xy}$ partial wave. 
We therefore believe that
much of the formula remains true even if more realistic wave functions
for the final states are chosen.\\
\begin{table}[htbp]
  \begin{center}
    \begin{tabular}{rr|rr|rrr|r}
      G$_x$/$2\pi$ & G$_y$/$2\pi$ & $\phi^{(opt)}_E$ &
      $I/I_0$&$\phi^{(opt)}_E
      $&$\theta^{(opt)}_E$&$I/I_0$& $h\nu$ \cr \hline               
   -1  & -1  &  2.356    &      1.908  &    2.356   &       1.571   &       1.908&22eV  \cr
   -1  &  0  &  1.275    &      0.958  &    1.275   &       1.005   &       1.345&22eV \cr
    0  & -1  &  0.295    &      0.958  &    0.295   &       1.005   &       1.345&22eV \cr
    0  &  0  &  2.356    &      1.000  &    2.356   &       1.571   &       1.000&22eV \cr \hline \hline
   -1  & -1  &  2.356    &      2.011  &    2.356   &       1.571   &       2.011&34eV \cr
   -1  &  0  &  0.974    &      1.053  &    0.974   &       1.040   &       1.417&34eV \cr
   -1  &  1  &  1.426    &      2.255  &    1.426   &       0.964   &       3.342&34eV \cr
    0  & -1  &  0.597    &      1.053  &    0.597   &       1.040   &       1.417&34eV \cr
    0  &  0  &  2.356    &      1.000  &    2.356   &       1.571   &       1.000&34eV \cr
    0  &  1  &  2.218    &      0.913  &    2.190   &       0.754   &       1.922&34eV \cr
    1  & -1  &  0.145    &      2.255  &    0.145   &       0.964   &       3.342&34eV \cr
    1  &  0  &  2.494    &      0.913  &    2.523   &       2.388   &       1.922&34eV \cr
    \end{tabular} 
    \caption{Intensities of the ZRS relative to $k=(\frac{\pi}{2},\frac{\pi}{2})$
      in higher Brillouin zones ${\bf k}+{\bf G}$ for 22 and 34eV photon energy.}
    \label{table1}
  \end{center}
\end{table}

\begin{table}[htbp]
  \begin{center}
    \begin{tabular}{rr|rr|rrr|r}
      G$_x$/$2\pi$ & G$_y$/$2\pi$ & $\phi^{(opt)}_E$ &
      $I/I_0$&$\phi^{(opt)}_E
      $&$\theta^{(opt)}_E$&$I/I_0$ & $h\nu$\cr \hline
   -1 &  -1  &    2.422   &       4.313 &   2.419   &       1.806    &      4.563&22eV \cr
   -1 &   0  &    0.000   &       1.000 &   0.000   &       0.556    &      3.436&22eV \cr
   -1 &   1  &    0.719   &       4.313 &   0.723   &       1.335    &      4.563&22eV \cr
    0 &  -1  &    0.719   &       4.313 &   0.723   &       1.806    &      4.563&22eV \cr
    0 &   0  &    0.000   &       1.000 &   0.000   &       2.586    &      3.436&22eV \cr
    0 &   1  &    2.422   &       4.313 &   2.419   &       1.335    &      4.563&22eV  \cr \hline \hline
   -2 &   0  &   0.000    &      1.645  &   0.000   &       2.318    &      3.058&34eV \cr
   -1 &  -1  &   2.472    &      2.588  &   2.466   &       1.963    &      3.019&34eV \cr
   -1 &   0  &   0.000    &      1.000  &   0.000   &       0.751    &      2.133&34eV \cr
   -1 &   1  &   0.669    &      2.588  &   0.675   &       1.178    &      3.019&34eV \cr
    0 &  -1  &   0.669    &      2.588  &   0.675   &       1.963    &      3.019&34eV \cr
    0 &   0  &   0.000    &      1.000  &   0.000   &       2.391    &      2.133&34eV \cr
    0 &   1  &   2.472    &      2.588  &   2.466   &       1.178    &      3.019&34eV \cr
    1 &   0  &   0.000    &      1.645  &   0.000   &       0.823    &      3.058&34eV \cr
    \end{tabular} 
    \caption{Intensities of the ZRS relative to $k=(\pi,0)$
      in higher Brillouin zones ${\bf k}+{\bf G}$ for 22 and 34eV photon energy.}
    \label{table2}
  \end{center}
\end{table}
\end{document}